\documentclass[preprint,amsmath,amssymb,pof,aip]{revtex4-1}

\usepackage{graphicx}
\usepackage{dcolumn}
\usepackage{bm}
\usepackage{color}

\begin{document}

\title{Analysis of granular rheology in a quasi-two-dimensional slow flow by means of discrete element method based simulations}
\author{Ashish Bhateja}
\email{ashish@iitgoa.ac.in}
\affiliation{School of Mechanical Sciences, Indian Institute of Technology Goa, Ponda 403401, Goa, India}
\author{Devang V. Khakhar}
\email{khakhar@iitb.ac.in}
\affiliation{Department of Chemical Engineering, Indian Institute of Technology Bombay, Powai, Mumbai 400076, India}
\date{\today}
\begin{abstract}
The steady flow of spherical particles in a rectangular bin is studied using the Discrete Element Method (DEM) for different flow rates of the particles from the bin, in the slow flow regime. The flow has two non-zero velocity components and is more complex than the widely studied unidirectional shear flows. The objective of the study is to characterize, in detail, the local rheology of the flowing material. The flow is shown to be nearly constant density, with a symmetric stress tensor and the principal directions of the stress and rate of strain tensors nearly colinear. The local rheology is analyzed using a coordinate transformation which enables direct computation of the viscosity and components of the pressure assuming the granular material to be a generalized Newtonian fluid. The scaled viscosity, fluctuation velocity and volume fraction are shown to follow power law relations with the \textit{inertial number}, a scaled shear rate, and data for different flow rates collapse to a single curve in each case. Results for flow of the particles on an inclined surface, presented for comparison, are similar to those for the bin flow, but with a lower viscosity and a higher solid fraction due to layering of the particles. The in plane normal stresses are nearly equal and slightly larger than the third component. All three normal stresses correlate well with the corresponding fluctuation velocity components. Based on the empirical correlations obtained, a continuum model is presented for computation of granular flows.
\end{abstract}
\maketitle
\section{Introduction}
Granular flows occur widely in industrial and natural systems \cite{jaeger1996,pouliquen_book}. Examples include mixing and segregation of granular mixtures\cite{ottino2000}, transport of particulates on conveyor belts\cite{cordero2015,zhu2019}, hopper discharge\cite{janda2012,peralta2017} in industry and debris flows\cite{takahashi1981}, snow and rock avalanches\cite{ancey2007} in geophysical settings. In most cases, the system geometry is complex, hence, computational modelling of the flows is required. Currently, the discrete element method (DEM) \cite{cundall1979,poschel,thornton,bkm2003,guo2015} is being extensively used since good predictions are obtained if particle properties and inter-particle forces are accurately incorporated in the simulations. The method provides access to particle-level information, which is often infeasible to obtain in experiments, and hence DEM has emerged as a powerful and reliable tool for studying the mechanics of granular media in the last four decades\cite{bkm2003,poschel,guo2015,thornton,bhateja2017}. However, discrete element computations are subject to the limitation of large computation times for systems with a large number of particles, which is typical in industrial and geophysical flows\cite{cleary2002,cleary2004,cleary2010}. Continuum models of granular flows are more suitable than DEM for such large systems. In addition to conservation equations, continuum models require equations to describe granular material behaviour\cite{forterre2018}. Although some constitutive models may be derived from first principles, most of the practically useful models have an empirical component and the model parameters of such constitutive equations need to be determined from experiments\cite{forterre2008}. This is the case for many complex fluids for which constitutive equations are incorporated into computational fluid dynamics codes for analysis of complex flow problems\cite{larson}.

Granular rheology, which has the objective of developing constitutive models for flow of particles, has been the subject of in-depth study for more than six decades\cite{bagnold1954,haff1983,jenkins1983,lun1984,pouliquen1999,silbert2001,midi2004,pouliquen2004,dacruz2005,jop2006,pouliquen2009,staron2010,tripathi2011,kamrin2012,bouzid2013,bouzid2015,seguin2016,mandal2016,kamrin2017,mandal2017,de2017,mandal2018,bharathraj2018,bhateja2018,berzi2018,sandip-pof2019,guo2016}, with the pioneering work of R. A. Bagnold\cite{bagnold1954} being an important initial milestone. Today, it is reasonably well-understood to the point that it is being employed for continuum simulations of many different kinds of granular flows\cite{staron2012,staron2014,martin2017,luo2019}. Kinetic theory applied to grains has provided the theoretical foundation for development of granular rheology\cite{haff1983,jenkins1983}. Using methods of dense gas kinetic theory, the balance laws for mass, linear momentum and fluctuation energy are derived for inelastic particles\cite{kumaran2015}. The equations are similar to those for non-isothermal flow of a Newtonian fluid, but with different constitutive relations, which incorporate the energy loss due to inelastic collisions 
\cite{jenkins1983,lun1984,jenkins1985,goldhirsch1996,berzi2015,saha2016,berzi2018,duan2019}. Several empirical approaches have also been suggested\cite{jop2006,kamrin2012,bouzid2015}. The $\mu$-$I$ model\cite{jop2006} is a notable recent development and is discussed below. It has the structure of generalized Newtonian fluid with a yield stress and has been found to give good predictions of the flow in a number of systems\cite{jop2006,lacaze2009,tripathi2011,yang2016,mandal2016,saingier2016,mandal2018,bhateja2018}. Barring a few exceptions\cite{lacaze2009,bhateja2018}, most of the studies to validate the $\mu$-$I$ model have been done for unidirectional shear flows. 

Non-zero normal stress differences, a deviation from Newtonian behaviour, have been observed in granular shear flows\cite{walton1986, campbell1986, jenkins1988, campbell1989, goldhirsch1996,alam2003,alam2005,saha2016}. Their origin is attributed to two primary factors: anisotropy in velocity fluctuations \cite{jenkins1988, saha2016} and microstructural changes due to inelastic collisions\cite{alam2003}. In a recent work, \citet{saha2016} showed that the normal stress differences for frictionless, inelastic particles in a simple shear flow are predicted by granular kinetic theory when second order Burnett terms are incorporated. The main contribution to the normal stress differences is due to anisotropy in the streaming normal stress components, arising from anisotropic velocity fluctuations\cite{saha2016}.

The objective of the present work is to obtain an empirical constitutive model for slow flow of granular material based on detailed analysis of data, with low error limits, generated by DEM simulations. A relatively complex steady flow in three dimensions, with two non-zero velocity components, is considered in conjunction with a transformation to enable unambiguous characterisation of the local rheology. Data are also presented for rheology of unidirectional shear flow (one non-zero velocity component) for comparison. The paper is organized as follows. Sec.~\ref{sec:simulation} provides details of the computational set-up and procedure followed for conducting the simulations presented in this study. The results are discussed in Sec.~\ref{sec:rd}, followed by conclusions in Sec.~\ref{sec:conclusion}.

\section{Simulation details}
\label{sec:simulation}
\begin{figure}[ht!]
\centering
\includegraphics[scale=0.22]{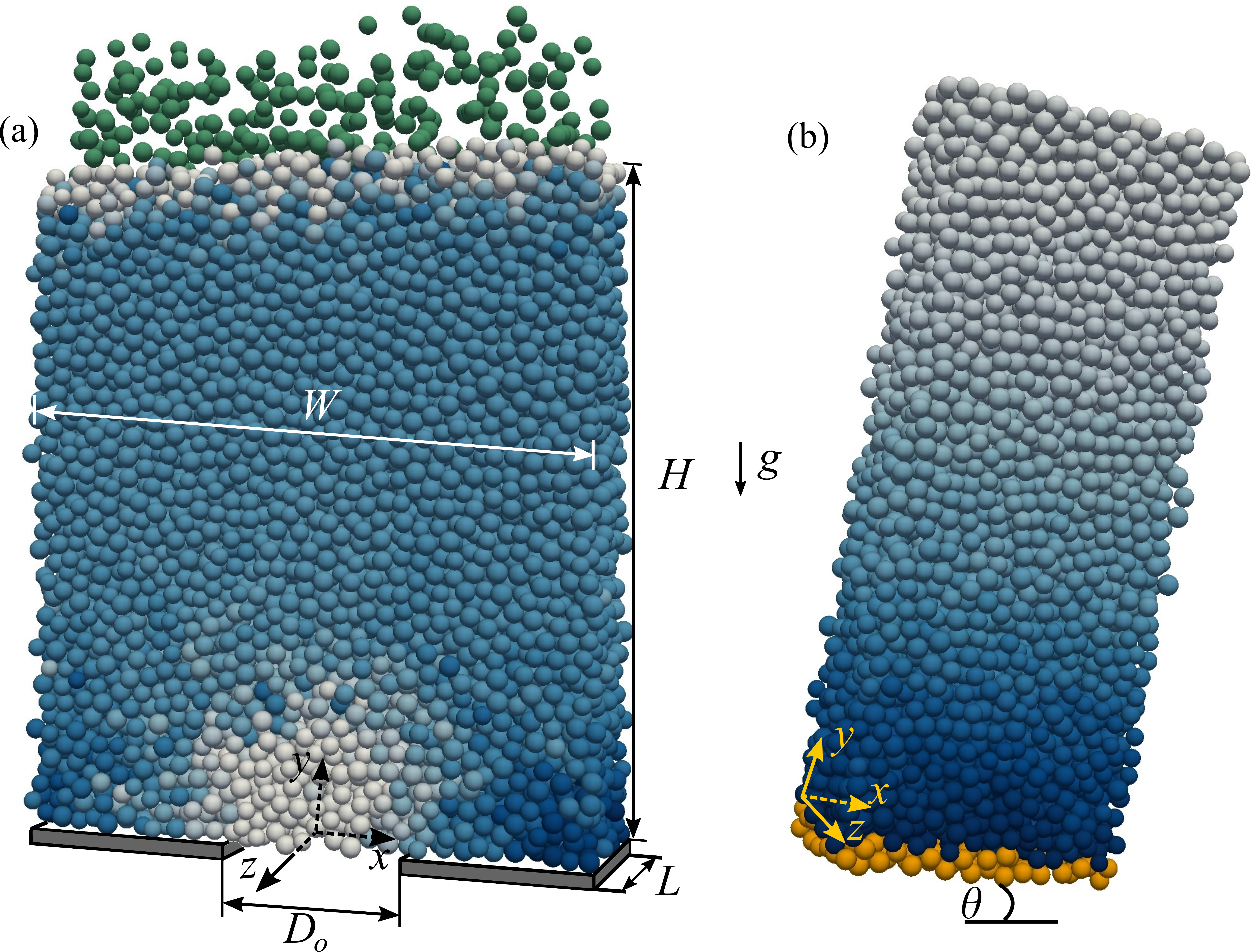}
\caption{(a) Gravity-induced flow of particles in a rectangular bin. Green particles shown at top are reinserted particles. (b) Particles flowing down a bumpy inclined surface (yellow particles). The direction of gravitational acceleration and coordinate axes are suitably displayed. Colours of the grains indicate speed, varying from dark blue to white with increasing speed.}
\label{fig:snapshots}
\end{figure}

Computations are carried out by means of the soft-particle discrete element method to simulate the flow of particles in a rectangular bin and over a bumpy inclined surface; the typical snapshots of both systems are shown in Fig.~\ref{fig:snapshots}. 

A flat frictional wall is used as the base for the bin and periodic boundary conditions are applied along all four sides. The width ($W$) and thickness ($L$) of the bin are taken to be $30d$ and $8d$, respectively, where $d$ is the mean diameter of the particles. The particles exit from a rectangular outlet, located at the center of the base of the bin, of width $D_o$ and spanning the thickness of the bin (Fig.~\ref{fig:snapshots}(a)). Three different outlet widths ($6d$, $7d$ and $8d$) are used, all larger than $5d$ in order to avoid arch formation and thus ensure a continuous flow of particles through the outlet\cite{zuriguel2005,mankoc2007}.  The number of particles used is 10,000,  corresponding to an initial fill height $H\simeq 36d$. The particles leaving the bin from the outlet are reinserted with zero velocity at random horizontal locations at heights $1d-6d$ above the free surface, as shown in Fig.~\ref{fig:snapshots}(a), thereby maintaining the fill height. We have confirmed that the flow characteristics are independent of $L$ by considering the systems with $L=12d$ and $16d$. We consider periodic boundary conditions along the $x$ and $z$ directions.

The inclined surface flow is simulated on a bumpy base comprising monodisperse spherical particles of diameter $d=1$, which is constructed by taking a slice of $1.2d$ thickness from a randomly closed packed configuration (see Fig.~\ref{fig:snapshots}(b)). The length and width along the flow direction ($x$) and vorticity direction ($z$) are $15d$ and $8d$, respectively. Periodic boundary conditions are applied along the $x$ and $z$ directions. We verified that the results are independent of system size by considering an inclined surface flow with twice of the width along $z$ direction.  5000 particles are used in the inclined surface flow simulations, which yields a flowing layer height approximately equal to that for the bin flow ($36d$). The angle $\theta$ of inclined surface is varied from $19.5^\circ$ to $22.5^\circ$, with an increment of $0.5^\circ$. 

The particles are considered to be inelastic and frictional spheres with a mass density $\rho$. A size polydispersity of $\pm 10\%$ is incorporated around mean diameter $d$, in order to prevent ordering in the system. The contact force between particles is modelled as a linear spring-dashpot along with a frictional slider\cite{shafer1996,zhang1996,stevens2005}. The normal component of the contact force comprises spring and viscous damping forces, whereas only a dashpot is employed for computing the tangential component, i.e., tangential spring stiffness $k_t=0$. The normal, $\bm{f}^n_{ij}$, and tangential, $\bm{f}^t_{ij}$, components of the force $\bm{f}_{ij}$ acting between two interacting particles $i$ and $j$ are given by 
\begin{eqnarray}
\bm{f}^n_{ij} &=& -k_n \, \delta_n \, \hat{\bm{n}}_{ij} - \gamma_n \, \bm{c}^n_{ij}, \\
\bm{f}^t_{ij} &=& - \gamma_t \, \bm{c}^t_{ij},
\end{eqnarray}
where $k_n$ and $\gamma_n$ are the spring stiffness and damping constant along normal direction, respectively, and $\gamma_t$ is the damping constant along tangential direction. The overlap between two particles in the normal direction is given by $\delta_n = [(R_i + R_j) - |\bm{r}_{ij}|]$, where $R_i$ and $R_j$ are the radii of particles $i$ and $j$, respectively, and $|\bm{r}_{ij}|$ is the distance between their centers. Here, $\bm{r}_{ij}=\bm{r}_j - \bm{r}_i$, with $\bm{r}_i$ and $\bm{r}_j$ being the position vectors of particles $i$ and $j$, respectively, with reference to a fixed coordinate system. The unit vector $\hat{\bm{n}}_{ij}$ points from the center of particle $i$ to the center of particle $j$, and is given by $\hat{\bm{n}}_{ij}=\bm{r}_{ij}/|\bm{r}_{ij}|$. The relative velocity at the point of contact between particles $i$ and $j$ is $\bm{c}_{ij} = \bm{c}_{i} - \bm{c}_{j} + (R_{i} \bm{\omega}_{i} + R_{j} \bm{\omega}_{j}) \times \hat{\bm{n}}_{ij}$, where $\bm{\omega}_i$ and $\bm{c}_i$ are, respectively, the angular velocity and the velocity of the center of mass of particle $i$, and `$\times$' denotes the cross product. The normal and tangential components of $\bm{c}_{ij}$ are $\bm{c}_{ij}^n$ and $\bm{c}_{ij}^t$, respectively, where $\bm{c}^n_{ij}=\bm{c}_{ij} - (\bm{c}_{ij} \cdot \hat{\bm{n}}_{ij})\,\hat{\bm{n}}_{ij}$ and $\bm{c}^t_{ij}=\bm{c}_{ij} - \bm{c}^n_{ij}$. The friction is modelled following Coulomb's friction criterion\cite{kruggel2008a}, limiting the value of the tangential component of force to $\mu_p\,|\bm{f}^n_{ij}|$, as follows
\begin{equation}
\bm{f}^t_{ij} = -\text{min}(\mu_p \, |\bm{f}^n_{ij}|\, \hat{\bm{t}}_{ij}, \gamma_t \, \bm{c}^t_{ij}),
\end{equation}
where $\hat{\bm{t}}_{ij} = \bm{c}^t_{ij}/|\bm{c}^t_{ij}|$ and $\mu_p$ is the friction coefficient between interacting particles.  The wall-particle contacts are modelled in the same manner as for particle-particle interactions.

All quantities of interest are made dimensionless by using $d$, $\rho$ and gravitational acceleration $g$ as characteristic parameters. In this study, the dimensionless normal spring stiffness, the restitution and friction coefficients for particle-particle interaction are $k_n=10^6$, $e_p=0.9$ and $\mu_p=0.4$, respectively. The restitution coefficient is independent of the impact velocity for linear spring-dashpot model and relates to the damping coefficient and normal spring stiffness\cite{shafer1996,kruggel2007}. The damping coefficients are taken to be equal for normal and tangential directions, i.e., $\gamma_n=\gamma_t$. The values of normal spring stiffness, restitution and friction coefficients for wall-particle interactions are same as those for particle-particle contacts. The equations of motion are integrated by utilizing the velocity-Verlet\cite{allen1989,kruggel2008b} integration scheme with a time step equal to $\Delta t=10^{-4}$.

The data presented for bin flow and inclined surface flow are averaged in steady state over 100 and 20 simulations runs, respectively, with each simulation run for 1 million time steps starting with a new initial configuration. We employ a coarse-graining technique \cite{weinhart2013} for averaging, considering a Heaviside step function with coarse-grained width $w$ equal to mean diameter $d$. The components $v_i$ of mean translational velocity vector $\bm{v}$ at the bin center are computed as
\begin{equation}
v_i = \frac {1}{N_b} \sum_{j=1}^{N_b} c_{ji},
\end{equation}
where $i=\{x,y,z\}$, $c_{ji}$ is the $i^{th}$ component of the instantaneous velocity of particle $j$ and $N_b$ is the number of particles lying within $w/2$ distance from the bin center such that $|x_j-x_b|\leq w/2$, $|y_j-y_b|\leq w/2$ and $|z_j-z_b|\leq w/2$, and $\bm{r}_j$=($x_j$,$y_j$,$z_j$) and $\bm{r}_b$=($x_b$,$y_b$,$z_b$) are the coordinates of particle $j$ and bin center, respectively. The solid fraction ($\phi$) of particles is calculated as
\begin{equation}
\phi = \frac {1}{N_b} \sum_{j=1}^{N_b} V_j/V_b,
\end{equation}
where $V_j$ and $V_b$ are the particle volume and bin volume, respectively. The components of fluctuation velocity vector $\bm{u}$ are calculated as
\begin{equation}
u_i = \sqrt{\frac {1}{N_b} \sum_{j=1}^{N_b} (c_{ji}-v'_{ji})^2},
\end{equation}
where $v'_{ji}$ are the components of the mean velocity vector $\bm{v}'_j$ computed at the center of particle $j$ as, following Artoni and Richard\cite{artoni2015},
\begin{equation}
\bm{v}'_j = \bm{v} + (\bm{r}_j - \bm{r}_b)\cdot \nabla \bm{v},
\end{equation}
where $\nabla \bm{v}$ is the velocity gradient tensor at the bin center. The total stress tensor $\bm{\sigma}$ comprising collisional $\bm{\sigma}_c$ and streaming $\bm{\sigma}_s$ components is given as
\begin{equation}
\bm{\sigma} = \bm{\sigma}_c + \bm{\sigma}_s.
\end{equation}
The collisional stress tensor is computed as\cite{tripathi2010}
\begin{equation}
\bm{\sigma}_c =\frac{1}{V_b} \sum_{c=1}^{N_c} \bm{f}_{ij} \otimes \overline{\bm{r}}_{ij},
\end{equation}
where $\bm{f}_{ij}$ is the force exerted on particle $i$ due to contacting particle $j$ and $\overline{\bm{r}}_{ij}=\bm{r}_i-\bm{r}_j$, $N_c$ is the number of contacts, and $\otimes$ denotes the dyadic product. The contribution of collisional stress in a bin is considered based on the location $\bm{r}_c$ of the contact between particles $i$ and $j$, where $\bm{r}_c=\bm{r}_j+R_j\,\hat{\bm{n}}$, $R_j$ is radius of particle $j$ and $\hat{\bm{n}}=\overline{\bm{r}}_{ij}/|\overline{\bm{r}}_{ij}|$, such that $|x_c-x_b|\leq w/2$, $|y_c-y_b|\leq w/2$ and $|z_c-z_b|\leq w/2$ and $\bm{r}_c$=($x_c$,$y_c$,$z_c$). The streaming stress tensor is given by\cite{tripathi2010}
\begin{equation}
\bm{\sigma}_s =\frac{1}{V_b} \sum_{j=1}^{N_b} m_j \bm{C}_j\otimes \bm{C}_j,
\end{equation}
where $m_j$ is the mass of particle $j$ and $\bm{C}_j	=\bm{c}_j-\bm{v}'_j$ is its fluctuation velocity.
\section{Results and discussions}
\label{sec:rd}
\subsection{Flow characteristics}
We first present results to characterise the flow in the bin at steady state. The system takes a time duration of approximately $t=50$ to reach steady state, starting from particles at rest. The average mass flow rates of particles through the outlet in the steady state for $D_o=6, 7$ and $8$ are $\dot{m}=47.03$, $63.47$ and $81.77$, respectively, with standard errors\cite{altman2005} less than 0.06\% when averaged over 90 configurations.

\begin{figure*}[ht!]
\centering
\includegraphics[scale=0.535]{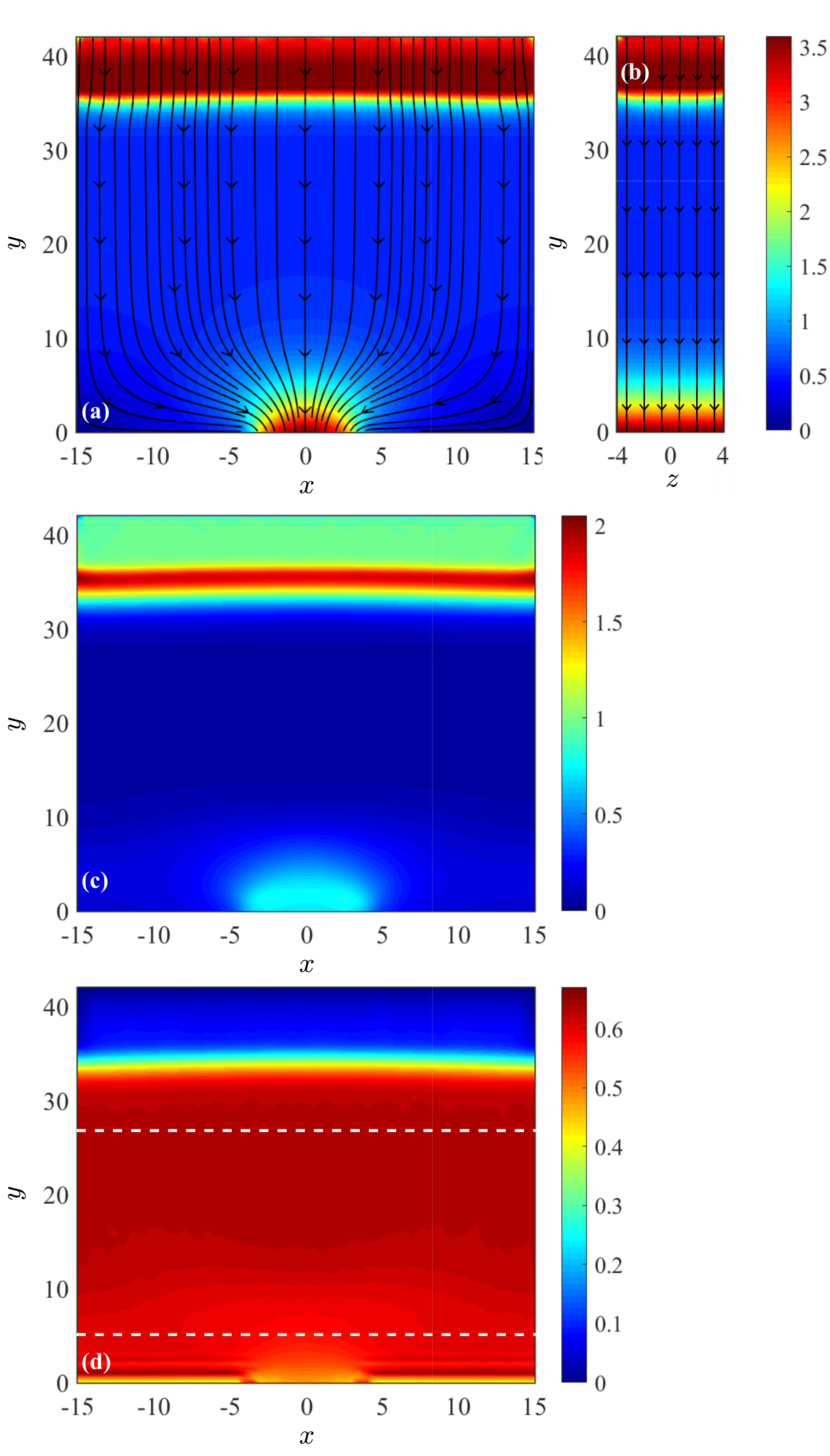}
\caption{(a) Streamlines and spatial distribution of the magnitude of mean velocity $v$. (b) Streamlines and spatial map of $v$ in y-z plane at $x=0$. Spatial distributions of (c) the magnitude of fluctuation velocity $u$, and (d) solid fraction $\phi$. The data are presented for $D_o=8$. Colour scales for all distributions are provided on their right side. The same color scale is used for (a) and (b), as shown on their right side.}
\label{fig:vpf}
\end{figure*}

Spatial maps of the magnitude of the mean velocity ($v$), magnitude of the fluctuation velocity ($u$) and the solid fraction ($\phi$) are presented in Fig.~\ref{fig:vpf} for outlet size $D_o=8$. Qualitatively similar plots are obtained for other outlet sizes. Fig.~\ref{fig:vpf}(a) shows the streamlines and spatial distribution of the velocity magnitude. As expected, the velocity increases towards the outlet and is nearly constant upstream of the flow. The higher velocities at the top are due to the impact of the freely falling reinserted particles on the free surface. The streamlines are nearly straight at the locations away from the outlet, indicating plug-like flow, and they converge into the outlet lower down. Further, the velocity does not vanish near the corners, indicating the absence of stagnant zones. This is because the flat frictional base allows particles to roll and slide. Fig.~\ref{fig:vpf}(b) shows spatial map of $v$ in $y-z$ plane at $x=0$. The velocity distribution is similar to what is described for Fig.~\ref{fig:vpf}(a). As expected, the streamlines  are straight, indicating that the flow is along $y$ direction. The behaviour of the fluctuation velocity ($u$) is similar to $v$, with low values in the upper part of the packed bed and increasing values near the outlet (Fig.~\ref{fig:vpf}(c)). The spatial distribution of solid fraction is shown in Fig.~\ref{fig:vpf}(d). The low values of $\phi$ in the top region correspond to the reinserted particles. The solid fraction is nearly constant and relatively high ($\phi \approx 0.6$) in the upper region of the bin and decreases in the vicinity of the outlet.

\begin{figure*}[ht!]
\centering
\includegraphics[scale=0.70]{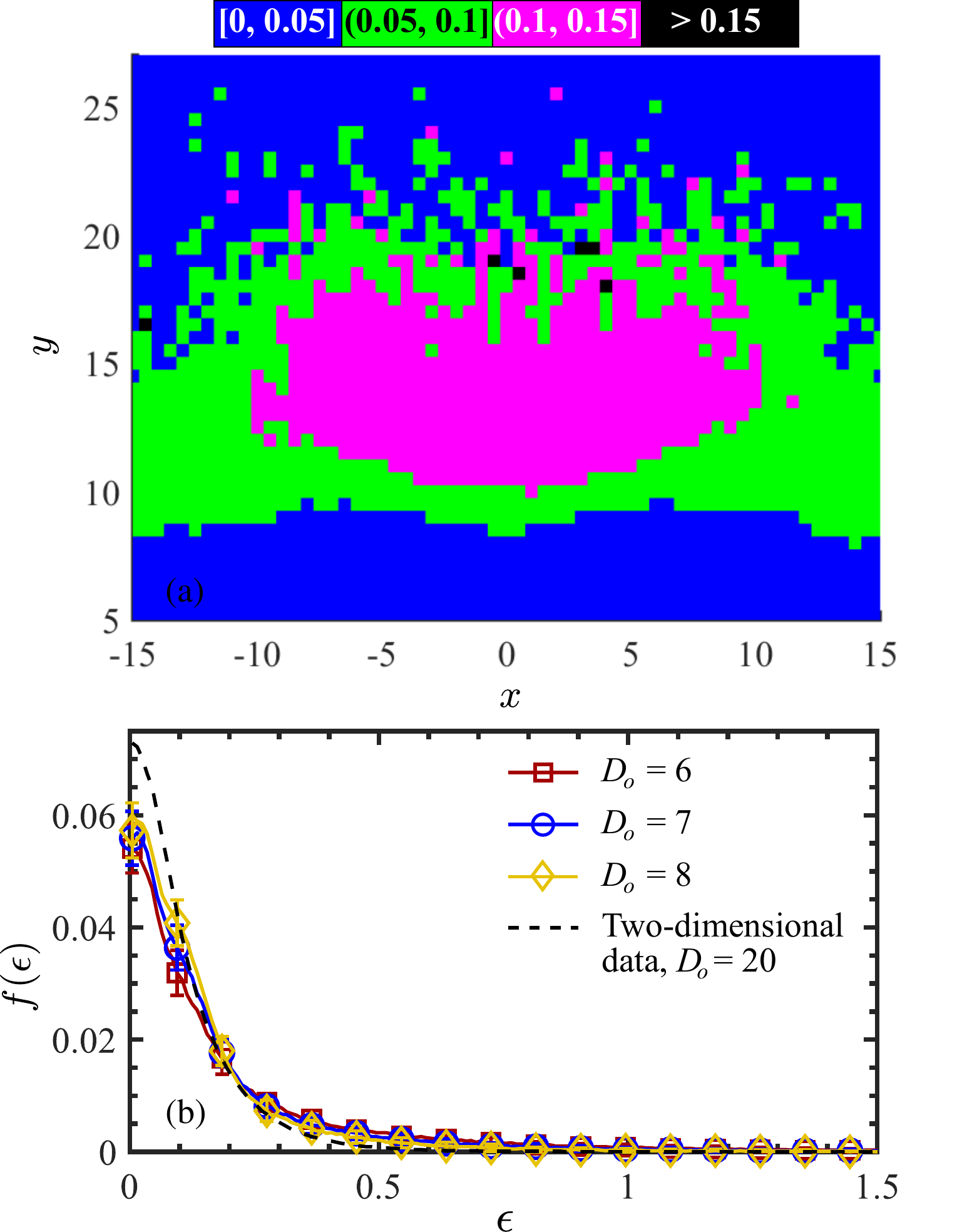}
\caption{(a) Spatial distribution of scaled dilation rate $\epsilon=|\nabla \cdot \bm{v}|/\dot{\gamma}$ for $D_o=8$. Qualitatively similar results are obtained for other $D_o$ values. The color scale for spatial distribution is given on its top. (b) The distribution function ($f({\epsilon})$) for all outlet sizes. For comparison, we show the data for one outlet size ($D_o=20$) for a two-dimensional  bin\cite{bhateja2018} (dashed line).}
\label{fig:SDR}
\end{figure*}

In order to avoid boundary effects, the results presented for the bin flow in the rest of this paper are for the region excluding a strip of thickness 5 particle diameters above the base and 9 particle diameters below the free surface, as indicated by the dashed lines in Fig.~\ref{fig:vpf}(d). Fig.~\ref{fig:SDR}(a) presents the spatial map of scaled dilation rate $\epsilon=|\nabla \cdot \bm{v}|/\dot{\gamma}$ for $D_o=8$, where $|\cdot|$ denotes the absolute value, $\bm{v}$ is the velocity vector and $\dot{\gamma}=\sqrt{2\bm{D}:\bm{D}}$ is the shear rate with $\bm{D}=(\bm{G}+\bm{G}^T)/2$ being the rate of deformation tensor given by the symmetric part of the traceless velocity gradient tensor, which is obtained from
\begin{equation}\label{eq:G}
\bm{G}=[\nabla \bm{v} - (\nabla \cdot \bm{v})\bm{I}/2]. 
\end{equation}
In most of the region, the dilation rate ($\nabla \cdot \bm{v}$) is less than $10\%$ of the shear rate (Fig.~\ref{fig:SDR}(a)). The distribution of the scaled dilation rate ($f(\epsilon)$), shown in  Fig.~\ref{fig:SDR}(b), also indicates that the dilation is small relative to the shear rate over most of the domain and the maximum value of the distribution function is nearly same for all outlet widths $D_o$.  Slightly higher values of $\epsilon$ are obtained here than earlier results in two dimensions\cite{bhateja2018}, shown as a dashed line in Fig.~\ref{fig:SDR}(b).  Results shown in Fig.~\ref{fig:SDR}(b) indicate that the flow is nearly isochoric and may be approximated to be incompressible.

\begin{figure*}[ht!]
\centering
\includegraphics[scale=0.70]{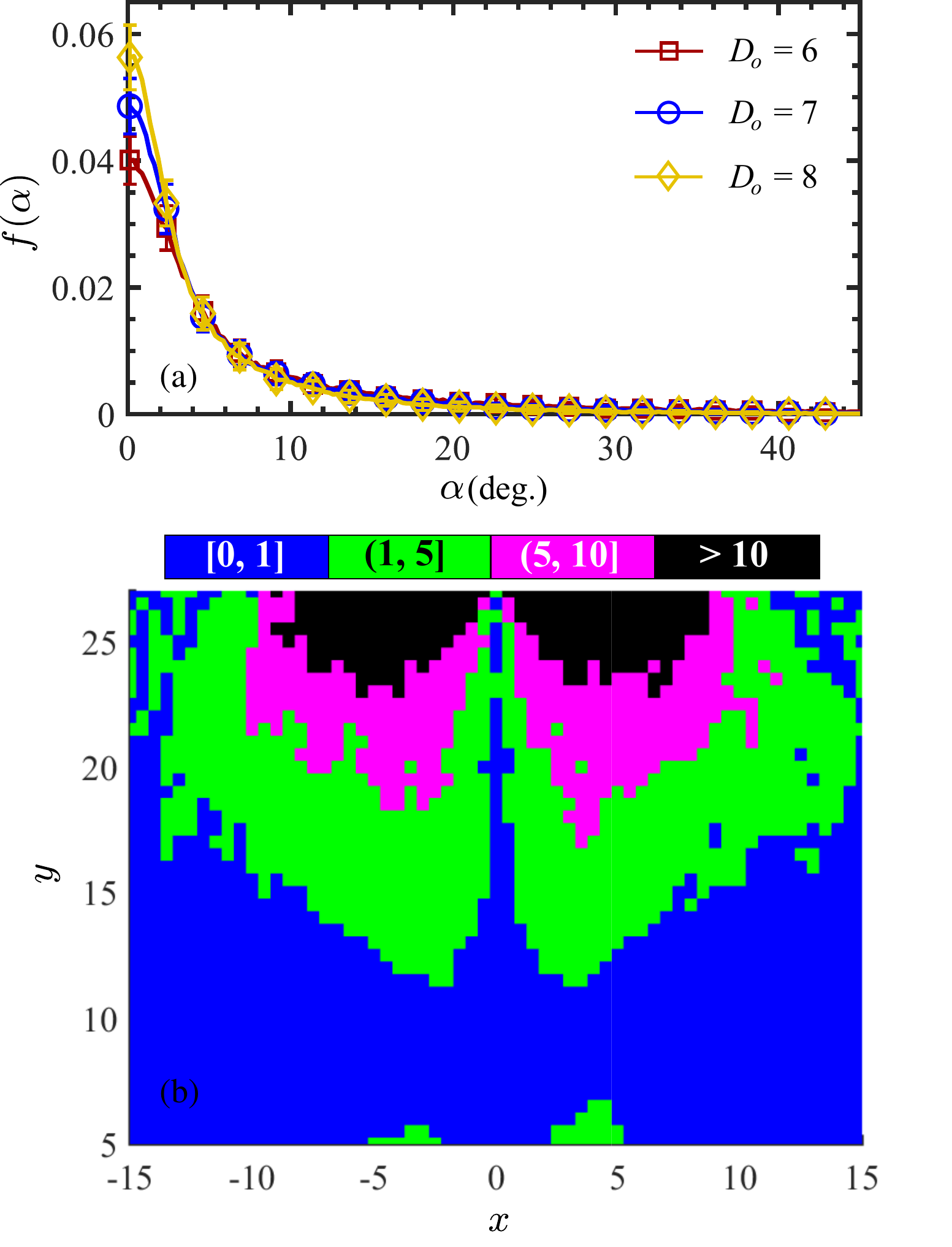}
\caption{(a) The distribution function $f(\alpha)$ for all outlet widths. (b) Spatial distribution of angle $\alpha$ for $D_o=8$. Qualitatively similar results are obtained for other $D_o$ values. The color scale for spatial distribution is given on its top.}
\label{fig:alpha}
\end{figure*}

Fig.~\ref{fig:alpha}(a) shows the distribution function $f(\alpha)$ for the angle $\alpha$ between the principal directions of the rate of deformation tensor $\bm{D}$ and deviatoric stress tensor $\bm{\tau}=-\bm{\sigma}+P\bm{I}$, where $P=\mbox{tr}(\bm{\sigma})/3$ is the pressure. The angle $\alpha$ is calculated by considering the absolute value of the dot product between the eigenvectors of $\bm{\tau}$ and $\bm{D}$, corresponding to the eigenvalues having the same sign. The peak occurs close to zero and the peak value of the distribution function rises as the outlet size becomes larger (Fig.~\ref{fig:alpha}(a)), indicating that the likelihood of alignment between tensors $\bm{\tau}$ and $\bm{D}$ increases with flow rate. The spatial distribution in Fig.~\ref{fig:alpha}(b) for $D_o=8$ confirms that $\alpha$ is close to zero in most of the region, implying the existence of colinearity, in agreement with the findings of Rycroft \textit{et al.}\cite{rycroft2009} for a geometrically similar system.

%

\begin{figure}[ht!]
\centering
\includegraphics[scale=0.7]{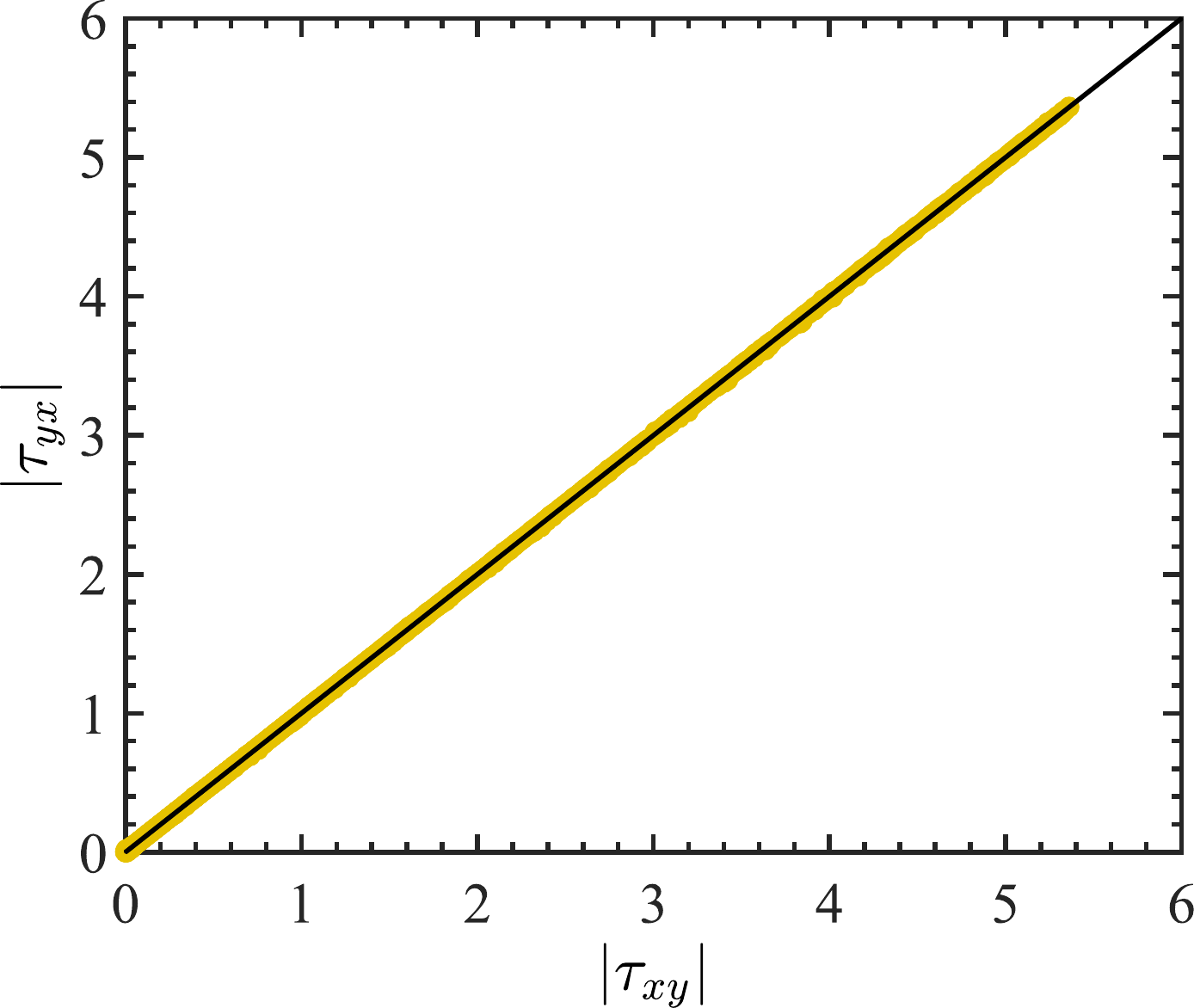}
\caption{Variation of stress component $|\tau_{yx}|$ with $|\tau_{xy}|$, where $|\cdot|$ denotes the absolute value. The data are plotted for $D_o=8$. Similar results are obtained for other outlet widths.}
\label{fig:ss}
\end{figure}

In order to ascertain symmetry of the stress tensor $\bm{\sigma}$, the stress components $\tau_{xy}$ and $\tau_{yx}$ are plotted for every point in the region of interest in Fig.~\ref{fig:ss}. All points lie on the diagonal $|\tau_{xy}|=|\tau_{yx}|$, confirming the symmetry of the stress tensor.

\subsection{Analysis of local rheology}
The results presented in the preceding section show that the stress tensor is symmetric to a high degree of accuracy. This implies that the angular momentum balance equation is identically satisfied\cite{rao_nott}. Further, the scaled dilation rate is small, indicating the flow may be approximated to be isochoric and the governing equations for the flow reduce to
\begin{equation}
\nabla\cdot {\bm v}=0,
\label{eqn:be1}
\end{equation}
\begin{equation}
\rho\phi{\bm v}\cdot\nabla{\bm v}=-\nabla\cdot{\bm \sigma}+\rho\phi {\bm g}.
\label{eqn:be2}
\end{equation}
In general, an additional equation for the balance of the kinetic energy of the velocity fluctuations ($u$) is required. Here, our approach is to seek scaling relations for $u$ so as to eliminate the need for the energy equation. 

The principal axes of the deviatoric part of the stress tensor ($\bm\tau$) and the rate of strain tensor ($\bm D$) are also shown to be nearly colinear, which is consistent with the granular material being a generalized Newtonian fluid, for which the total stress tensor ($\bm\sigma$) is given by
\begin{equation}\label{eqn:sig1}
{\bm \sigma}={\bm P}\cdot{\bm I} -{\bm \tau}={\bm P}\cdot{\bm I} -2\eta{\bm D}
\end{equation}
where $\eta$ is the viscosity and 
\begin{equation}
\bm{P}=\left(\begin{array}{ccc} P_x &0 &0 \\
 0 &P_y  &0\\
 0 &0 &P_z\end{array}\right)
\end{equation}
is the pressure tensor, taking into consideration that the components of the pressure may be different in magnitude.   We transform the local coordinate system, following \citet{bhateja2018}, to obtain the components of the pressure and the viscosity, assuming the material to be described by Eq.~(\ref{eqn:sig1}).

\begin{figure}[ht!]
\centering
\includegraphics[scale=0.4]{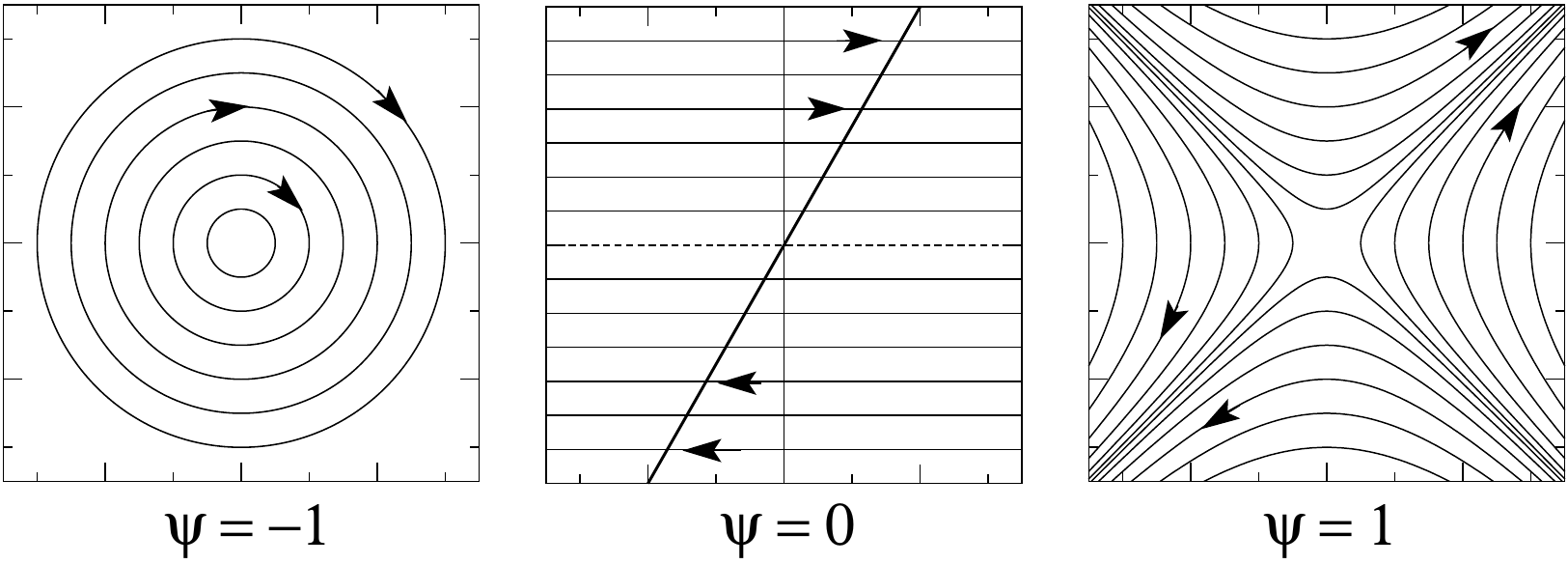}
\caption{A schematic representation of the streamlines of different kinds of flow.}
\label{fig:flow-type}
\end{figure}

As shown below, any two-dimensional isochoric flow can be linearized at a point and transformed to the following form
\begin{equation}
v_{x'}=\dot{\gamma}y',\qquad v_{y'}=\psi\dot{\gamma}x',\label{eq:2dflow}
\end{equation}
where $\dot{\gamma}$ is the shear rate and $(x',y')$ are the transformed coordinates. The parameter $\psi$ defines the local flow type\cite{wagner2016,lee2007,hudson2004}: $\psi = 0$ corresponding to simple shear flow, $\psi = 1$ corresponding to pure extensional flow and $\psi =-1$ to solid body rotation. Schematic views of the flows for different $\psi$ values are shown in Fig.~\ref{fig:flow-type}. The velocity gradient tensor for the flow is given by
\begin{equation}
\nabla{\bm v}=\left(\begin{array}{cc}0 &\dot{\gamma}  \\
 \psi\dot{\gamma} & 0 \end{array}\right).
\end{equation}
Since the diagonal components of the velocity gradient tensor are zero, the normal stress components for a generalized Newtonian fluid do not have a viscous contribution and are entirely due to the pressure. The stress tensor thus simplifies to
\begin{equation}
{\bm\sigma}=\left(\begin{array}{ccc} P_{x'} &\tau_{x'y'} &0 \\
 \tau_{x'y'} &P_{y'}  &0\\
 0 &0 &P_z\end{array}\right).
 \label{eqn:sigma}
\end{equation}
where  $\tau_{x'y'}$ is the shear stress and $P_{x'}$, $P_{y'}$ and $P_z$ are the components of the pressure. In the transformed coordinates, the relation between the shear stress and shear rate can be directly obtained as can the individual components of the pressure.

The transformation of the velocity gradient tensor is carried out as follows. The isochoric part of the velocity gradient tensor ($\bm{G}$) is obtained from Eq.~(\ref{eq:G}). Transforming ${\bm G}$ into a tensor with zeros as diagonal elements is equivalent to transforming ${\bm G}'= {\bm Q}\cdot{\bm G}$ into a diagonal tensor, where ${\bm Q} = [0\quad 1; - 1\quad 0]$ is the orthogonal rotation matrix. The latter transformation corresponds to finding the eigenvalues of ${\bm G}'$ as
\begin{equation}
{\bm G}'\cdot{\bm k}=\lambda{\bm k}
\end{equation}
where ${\bm k}$ is the eigenvector for eigenvalue $\lambda$. Taking into account the condition $\mbox{tr}({\bm G}) = 0$, the general form for $\bm G$  is
\begin{equation}
{\bm G}=\left(\begin{array}{cc}a &b  \\
 c &-a \end{array}\right),
\end{equation}
which yields the following eigenvalues of ${\bm G}'$
\begin{eqnarray}
\lambda_1&=&(c-b)/2+\left[(c+b)^2/4+a^2\right] ^{1/2}\label{eq:lam1}\\
\lambda_2&=&(c-b)/2-\left[(c+b)^2/4+a^2\right]^{1/2}.\label{eq:lam2}
\end{eqnarray}
Since the term in square brackets in Eqs.~(\ref{eq:lam1}) and (\ref{eq:lam2}) is always positive, real eigenvalues are obtained for all $\bm G$. This proves that every two-dimensional isochoric flow can be cast in the form given in Eq.~(\ref{eq:2dflow}). The transformed tensor is then
\begin{equation}
{\bm G}= {\bm Q}^T{\bm G}' = \left(\begin{array}{cc}0 &-\lambda_2  \\
 \lambda_1 &0 \end{array}\right),
\end{equation}
using the identity ${\bm Q}^T\cdot{\bm Q}={\bm I}$.

The basis vectors for the transformed coordinate system are ${\bm k}_1$, ${\bm k}_2$ and ${\bm k}_3$, where ${\bm k}_1$ and ${\bm k}_2$ are the eigenvectors, of unit magnitude, corresponding to eigenvalues $\lambda_1$ and $\lambda_2$, respectively, and ${\bm k}_3={\bm e}_z$ is the unit vector normal to ${\bm k}_1$ and ${\bm k}_2$, where ${\bm e}_z$ is the unit vector in the $z$-direction. Based on the above transformation, the viscosity is given by
\begin{equation}
\eta =\frac{ {\bm k}_1\cdot{\bm \sigma}\cdot{\bm k}_2}{{\bm k}_1\cdot{\bm D}\cdot{\bm k}_2},
\end{equation}
and the components of the pressure are given by
\begin{eqnarray}
P_{x'}&=&{\bm k}_1\cdot{\bm \sigma}\cdot{\bm k}_1,\\
P_{y'}&=&{\bm k}_2\cdot{\bm \sigma}\cdot{\bm k}_2,
\end{eqnarray}
with $P_z=\sigma_{zz}$.

\subsection{Scaling relations}
We consider here empirical scaling relations for the viscosity, volume fraction, fluctuation velocity and pressure based on the data for the bin flow in the transformed coordinates. The primes are omitted for brevity. Data for the inclined surface flow are shown for comparison; the results presented for the inclined surface flow are for the region lying $5$ particle diameters above the base and $6$ particle diameters below the free surface. Only data points with standard error\cite{altman2005} less than 2\% of the mean value are considered in the analysis, unless stated otherwise. By doing so, we ensure that the spread in the data is inherent and not because of the computational error. The objective is to obtain simple but accurate relations for the rheological parameters. A number of different empirical correlations were considered; we report here only those that gave the best results for our systems.

\begin{figure}[ht!]
\centering
\includegraphics[scale=0.7]{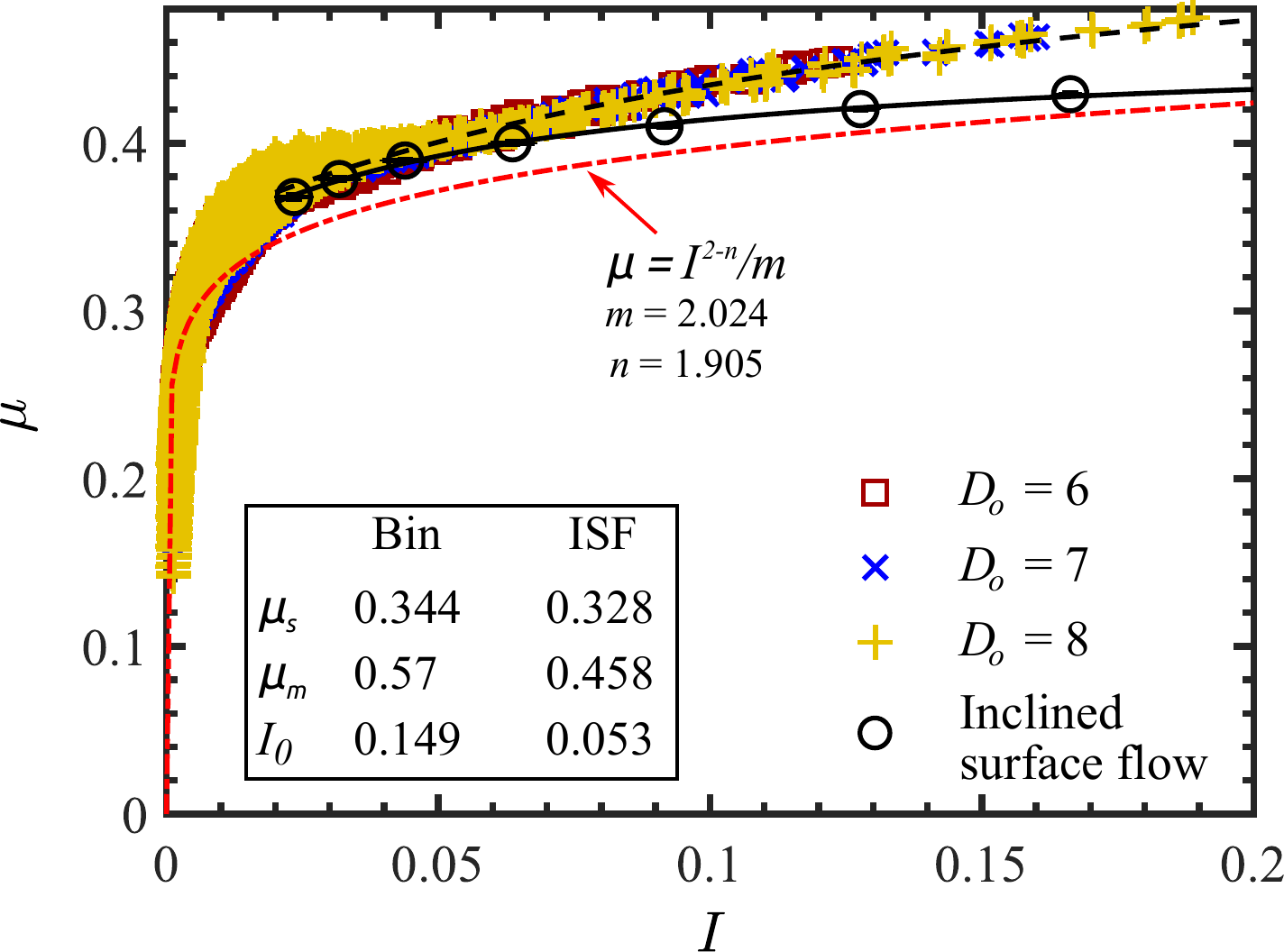}
\caption{Variation of the effective friction coefficient $\mu$ with inertial number $I$. Solid and dashed lines represent fits of Eq.~(\ref{eqn:jop}) for the inclined surface flow and bin flow, respectively. The fitted values of the model parameters are given in the inset for both systems.}
\label{fig:mu}
\end{figure}

Consider first a comparison of the simulation data to the model of Jop \textit{et al.}\cite{jop2006}, for which the viscosity is given by
\begin{equation}
\eta = \mu P/\dot{\gamma},
\end{equation}
where $\mu$ is the friction coefficient defined as
\begin{equation}
\mu=|\tau_{xy}|/P,
\end{equation}
with the shear rate given by $\dot{\gamma}=(2\bm{D:D})^{1/2}=2|D_{xy}|$ and the pressure by $P=(P_x+P_y+P_z)/3$. The above definitions are in terms of the transformed coordinates. The variation of the friction coefficient ($\mu$) with the inertial number ($I$) is shown in Fig.~\ref{fig:mu} for both bin flow and inclined surface flow, where the inertial number, a scaled shear rate, is defined as
\begin{equation}
I=\dot{\gamma}d/(P/\rho)^{1/2}.
\end{equation}
The data for bin flow for the different outlet sizes collapse reasonably well to a single curve, as found previously for different systems\cite{jop2006,lacaze2009,tripathi2011,mandal2016}.  The spread in the data for the bin flow is larger than the standard error. Two regions can be identified for the bin flow data: a region of steep rise in friction coefficient, from $\mu \approx 0.15$ to $\mu \approx 0.35$, at low $I$ values  ($I<0.02$) and a gradual rise thereafter. The data for $I>0.02$ are well-described by the relation proposed by Jop \textit{et al.}\cite{jop2006} (dashed line in Fig.~\ref{fig:mu}),
\begin{equation}
\mu = \mu_s + \frac{\mu_m-\mu_s}{(1+I_0/I)},
\label{eqn:jop}
\end{equation} 
where $\mu_s$, $\mu_m$ and $I_0$ are model parameters. The inclined surface flow data are also well described by Eq.~(\ref{eqn:jop}) (solid line in Fig.~\ref{fig:mu}), but the data for the inclined surface flow do not fall on those for bin flow, as reported by us for the case of two-dimensional systems\cite{bhateja2018}. Fitted values of the model parameters for both flows are given in the inset of Fig.~\ref{fig:mu}. The values obtained for $I_0$ are lower as compared to those reported earlier, but $\mu_s$ and $\mu_m$ are reasonably close to the reported values \cite{jop2006,tripathi2011}. The flows span relatively low values of the inertial number ($I<0.15$) and correspond to slow, dense flows. The data at very low values of inertial number ($I<0.02$) for bin flow indicate complex yielding behaviour and $\mu$ values significantly smaller than $\mu_s$ are obtained at low $I$.

\begin{figure}[ht!]
\centering
\includegraphics[scale=0.6]{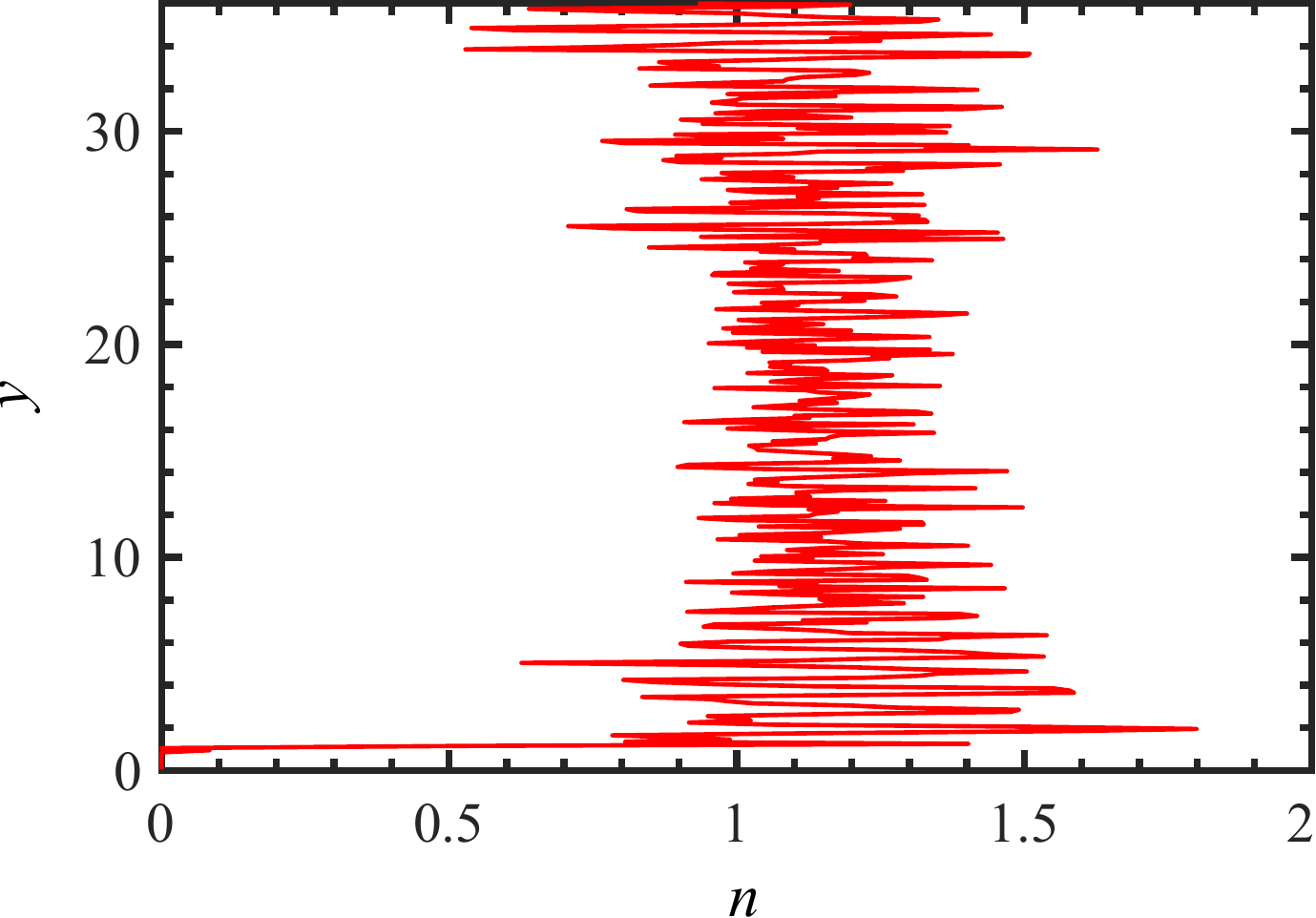}
\caption{Number density ($n$) for inclined surface flow at $\theta=22.5^{\circ}$. Qualitatively similar behaviour is obtained for other $\theta$ values.}
\label{fig:layering}
\end{figure}

The difference in the data for bin flow and inclined surface flow is not directly related to the flow type parameter ($\psi$), as shown previously for two-dimensional systems\cite{bhateja2018}. Fig.~\ref{fig:layering} shows the variation of number density ($n$) with height ($y$) for the inclined surface flow; the data are averaged over $200$ configurations sampled at a time interval of $t=0.01$. The peaks in the graph are separated by a distance of one diameter indicating layering of particles, in spite of the bumpy base. No such layering is seen in case of the bin flow. The lower values of the friction coefficient ($\mu$) for the inclined surface flow relative to the bin flow are due to this layering, which has been shown to reduce the viscosity/friction coefficient of the flow \cite{silbert2001,kumaran2012,mandal2017}. 

\begin{figure}[ht!]
\centering
\includegraphics[scale=0.65]{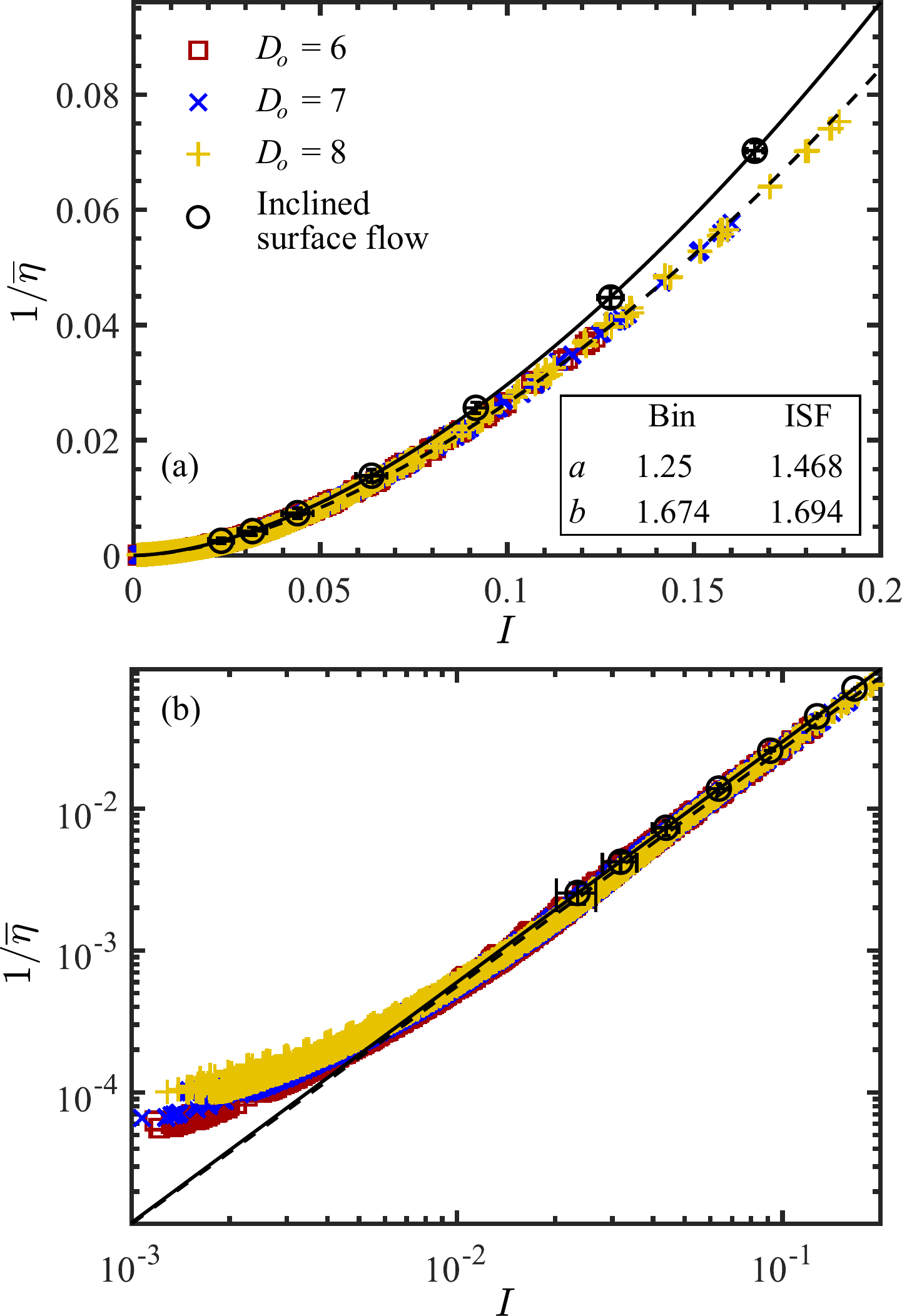}
\caption{Variation of the inverse of the scaled viscosity $1/\overline{\eta}$ with $I$ on (a) linear-linear scale and (b) log-log scale. Dashed and solid lines are fits of Eq.~(\ref{eqn:eta_I}) for the bin flow and inclined surface flow, respectively. Legend for both plots is given in (a).}
\label{fig:eta_fvel}
\end{figure}

Kinetic theories for granular flow yield a scaling of the form $\eta=\rho ud F(\phi)$ for the viscosity\cite{jenkins1985,campbell2006,kumaran2015}, where $u$ is the fluctuation velocity and $\phi$ is the solid fraction.  Fig.~\ref{fig:eta_fvel} shows the variation of the inverse of scaled viscosity ($1/\overline{\eta}$) with the inertial number $I$, where $\overline{\eta}=\eta/\rho u d$  and $\eta=\tau_{xy}/2D_{xy}$. The inverse of scaled viscosity is plotted in order to emphasize the lower values of viscosity and the inertial number ($I$) is used instead of the volume fraction ($\phi$) since a better collapse is obtained for $I$. The data for bin flow for all outlet sizes collapse very well to a single curve and the spread in the data is less than that obtained for the $\mu$-$I$ scaling. The data are fitted to a power law relation,
\begin{equation}
1/\overline{\eta} = a I^b,
\label{eqn:eta_I}
\end{equation}
and fitted values of the model parameters, $a$ and $b$, are given in the inset of Fig.~\ref{fig:eta_fvel}(a), considering data points with $I>0.01$. Fig.~\ref{fig:eta_fvel}(b) shows the data and fitted curve on a log scale, where the deviation of the data at low $I$ from the correlation is evident. The inclined surface flow shows a similar variation as bin flow but with a lower viscosity: the exponent in the power law ($b$) is nearly the same as that for bin flow but the value of the prefactor ($a$) is higher.

\begin{figure}[ht!]
\centering
\includegraphics[scale=0.7]{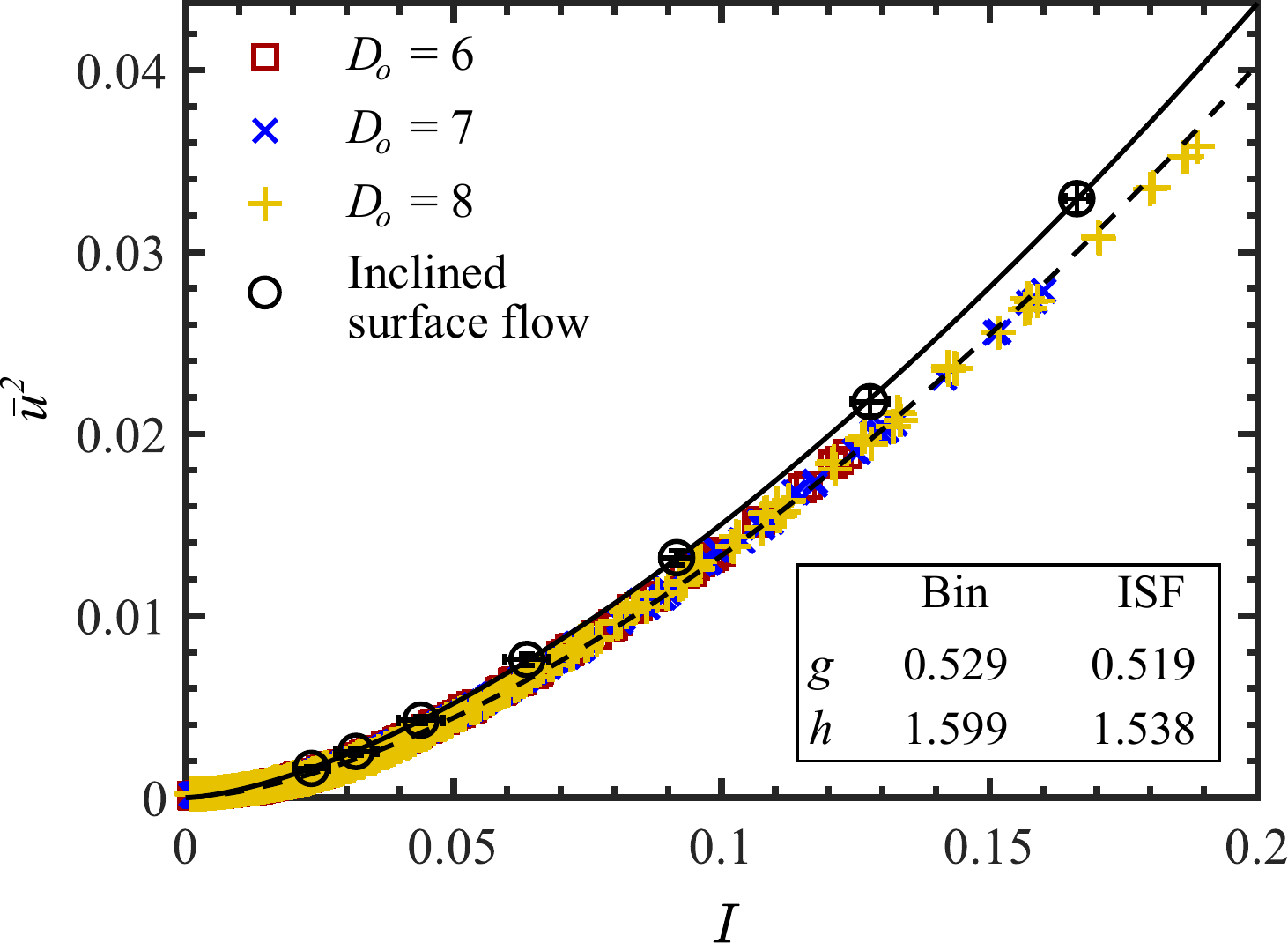}
\caption{Variation of the square of scaled fluctuation velocity $\overline{u}^2$ with $I$. Solid and dashed lines are fits of Eq.~(\ref{eqn:PI}) considering inclined surface flow and bin flow, respectively.}
\label{fig:press_I}
\end{figure}

Scaling of viscosity with the fluctuation velocity ($u$) introduces an additional variable compared to the $\mu$-$I$ scaling. Use of the scaling would thus require an additional equation for closure, as mentioned above. One possibility is to solve the energy balance equation \cite{kumaran2015} along with Eqs.~(\ref{eqn:be1}) and (\ref{eqn:be2}) to obtain the fluctuation velocity field, which would increase the computational difficulty of the problem. Here, we consider obtaining an empirical relation for $u$.  Kinetic theory indicates that the pressure scales as $P = \rho u^2 f(\phi)$\cite{kumaran2015}. We thus define the scaled fluctuation velocity as $\overline{u}=u/(P/\rho)^{1/2}$. The variation of $\overline{u}^2$ with $I$ is shown in Fig.~\ref{fig:press_I} for both systems. Again, the data for all outlet sizes for the bin flow collapse very well to a single curve, while higher values are obtained for the inclined surface flow. A power-law expression fits the data for both systems
\begin{equation}
\overline{u}^2=g I^h,
\label{eqn:PI}
\end{equation}
and the fitted model parameters $g$ and $h$ are given in the inset of Fig.~\ref{fig:press_I} considering points with $I>0.01$ for the fitting.

%

\begin{figure}[ht!]
\centering
\includegraphics[scale=0.6]{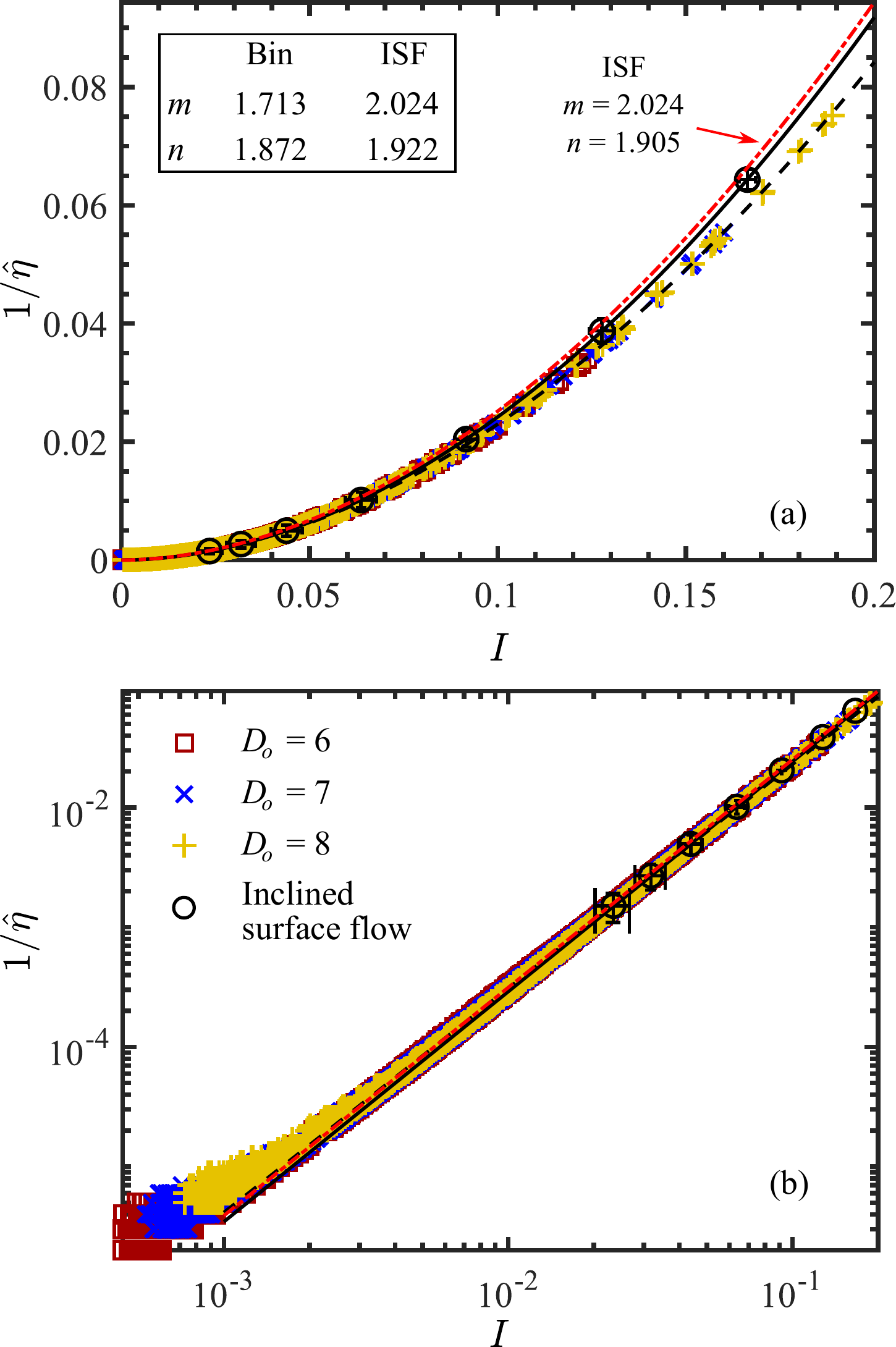}
\caption{Variation of the inverse of the scaled viscosity $1/\hat{\eta}$ with $I$ on (a) linear-linear scale and (b) log-log scale. Dashed and solid lines are fits of Eq.~(\ref{eqn:eta_I_1}) considering the bin flow and inclined surface flow, respectively. Legend for both plots is given in (b).}
\label{fig:eta_gdot}
\end{figure}

The data collapse obtained for $\overline{u}$ with $I$ in Fig.~\ref{fig:press_I} implies that  Eqs.~(\ref{eqn:eta_I}) and (\ref{eqn:PI}) can be combined to eliminate $\overline{u}$ for a scaling of the viscosity with the shear rate as $\hat{\eta}=\eta/(\rho \dot{\gamma}d^2)$, which varies with inertial number as
\begin{equation}
1/\hat{\eta}= m I^n,
\label{eqn:eta_I_1}
\end{equation}
where $m$ and $n$ are the model parameters. The variation of $1/\hat{\eta}$ with $I$ is shown in Fig.~\ref{fig:eta_gdot} for all outlet sizes for the bin flow and all inclinations for the inclined surface flow, considering the data points where standard error is less than 5\%. Here, the upper bound of standard error is relaxed as a sufficiently large number of points are not obtained while taking the upper limit to  be 2\%. Expectedly, the data collapse quite well for both systems, considering the data collapse for $\overline{\eta}$ and $\overline{u}$. The power law relation for the bin flow data extends to smaller values of inertial number than those for $\overline{\eta}$ (cf. Figs.~\ref{fig:eta_gdot}(b) and \ref{fig:eta_fvel}(b)), hence, data for $I>0.003$ are considered for fitting and the fitted model parameters for each system are given in the inset of Fig.~\ref{fig:eta_gdot}(a). The exponent $n$ is similar for the two systems, as also seen for the case of $\overline{\eta}$. In order to test Eq.~(\ref{eqn:eta_I_1}), we plot $x$-component of velocity ($v_x$) with vertical distance ($y$) for the inclined surface flow for four different angles $\theta$ as shown in Fig.~\ref{fig:xvel_fit}. Solid lines are fits of the following equation (see Appendix for derivation)
\begin{equation}
v_x = \dfrac{2}{3d} (m\, \text{tan}\theta)^{1/(2-n)} (g \phi \text{cos}\theta)^{1/2} \left\lbrace H^{3/2}-(H-y)^{3/2} \right\rbrace.
\label{eqn:xvel_fit}
\end{equation}
For $m=2.024$ and $n=1.922$, the prediction of Eq.~(\ref{eqn:xvel_fit}) differs substantially from the simulation data. However, a good match is obtained between them by tuning the exponent $n$ to $1.905$ as shown in Fig.~\ref{fig:xvel_fit}. It is reasonable to consider $n=1.905$ as there is a small difference between two curves for the inclined surface flow when $1/\hat{\eta}$ is plotted by considering $n=1.905$ (solid and dash-dotted lines in Fig.~\ref{fig:eta_gdot}).

Of the three cases considered, i.e., $\mu,\overline{\eta}$ and $\hat{\eta}$, the viscosity scaled with the shear rate ($\hat{\eta}$) gives the best collapse of data. It is noteworthy that $\hat{\eta}$ can be expressed in terms of $\mu$ and $I$ as follows, 
\begin{equation}
\hat{\eta} = \dfrac{\eta}{\rho \dot{\gamma} d^2} = \dfrac{\mu P}{\rho \dot{\gamma}^2 d^2} = \dfrac{\mu}{I^2},
\label{eqn:eta_hat_mu}
\end{equation}
where $\eta = \mu P/\dot{\gamma}$ according to the model of Jop \textit{et al}. Considering Eqs.~(\ref{eqn:eta_I_1}) and (\ref{eqn:eta_hat_mu}), we get
\begin{equation}
\mu = \dfrac{I^{2-n}}{m}.
\label{eqn:mu_I_2}
\end{equation}
The dash-dotted line in Fig.~\ref{fig:mu} represents Eq.~(\ref{eqn:mu_I_2}) for $m=2.024$ and $n=1.905$, predicting lower values of $\mu$ in comparison with what Eq.~(\ref{eqn:jop}) does. It is worth noting that Eq.~(\ref{eqn:mu_I_2}) differs from Eq.~(\ref{eqn:jop}) in the sense that in case of the former $\mu$ vanishes when $I$ decreases to zero, however, the latter reduces $\mu$ to a finite value. We note that $1/\hat{\eta}$ corresponds to a local Reynolds number based on the shear rate, $Re_G=\rho\dot{\gamma}d^2/\eta$. The data in Fig.~\ref{fig:eta_gdot} show that $Re_G<0.07$ indicating the flow is in the creeping flow regime. Computations of the macroscopic Reynolds number ($Re=\rho v D_o/\eta$) over the domain gave $Re<1$ everywhere, again indicating that viscous effects dominate.

\begin{figure}[ht!]
\centering
\includegraphics[scale=0.7]{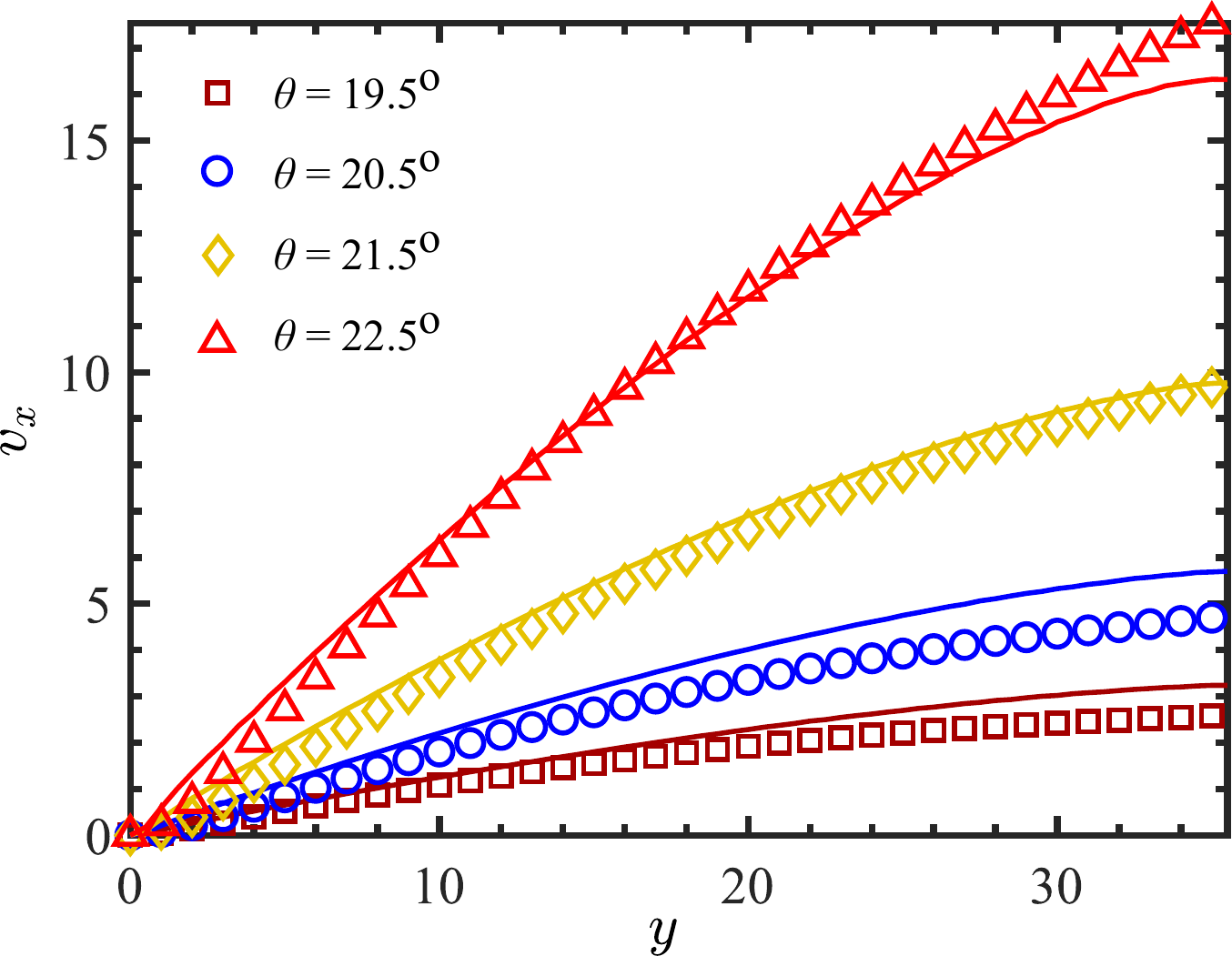}
\caption{Variation of $x$-component of velocity ($v_x$) for the inclined surface flow with vertical distance ($y$) from the bumpy base. Solid lines are fits of Eq.~(\ref{eqn:xvel_fit}) for $m=2.024$ and $n=1.905$.}
\label{fig:xvel_fit}
\end{figure}
%

\begin{figure}[ht!]
\centering
\includegraphics[scale=0.7]{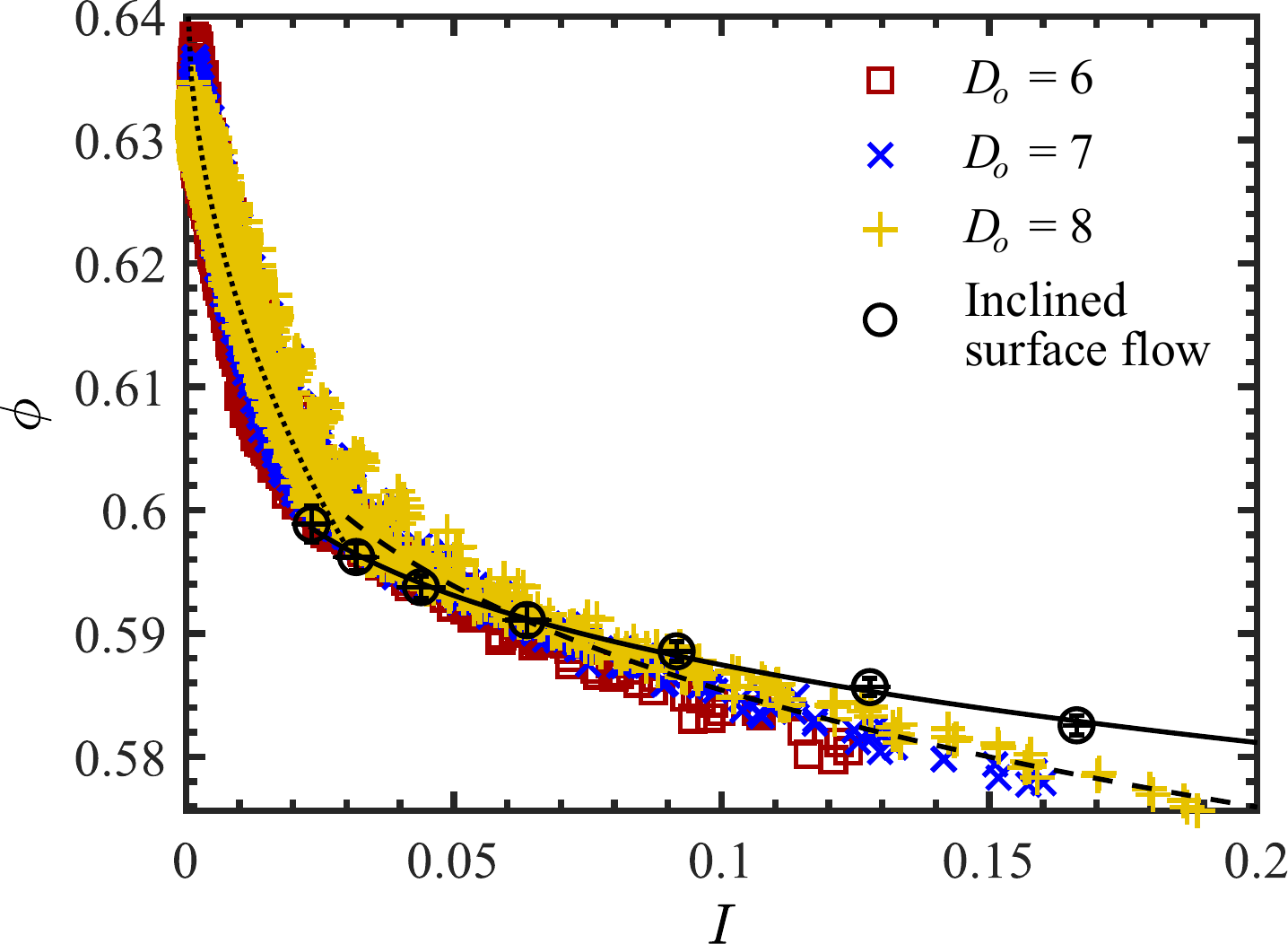}
\caption{Variation of solid fraction $\phi$ as a function of inertial number $I$. Solid (inclined surface flow), dotted (bin flow, $I\leq 0.03$) and dashed (bin flow, $I>0.03$) lines are fits of Eq.~(\ref{eqn:phi}).}
\label{fig:phi}
\end{figure}

Fig.~\ref{fig:phi} shows the variation of solid fraction $\phi$ with the inertial number $I$ for both systems. Slightly higher volume fractions are obtained for the inclined surface flow than the bin flow for a given value of the inertial number ($I$); this is again due to layering of the particles in the case of the inclined surface flow\cite{silbert2001,kumaran2012,mandal2017}. There is a reasonable collapse of the bin flow data for different outlet sizes to a single curve, though the scatter in this case is larger.  Similar to the $\mu$-$I$ data, two regions of $\phi$-$I$ variation can be seen: a region of rapid decrease in solid fraction from $\phi \approx 0.64$ to $\phi \approx0.60$ for small inertial numbers ($I \leq 0.03$) and a gradual decrease in $\phi$ with $I$ for $I>0.03$. We fit the following power-law relation to data for both systems\cite{hatano2007},
\begin{equation}
\phi = \phi_m - rI^s, 
\label{eqn:phi}
\end{equation}
with the bin flow data fitted in two inertial number ranges, i.e., $I\leq0.03$ and $I> 0.03$; $\phi_m$. The fitted values of the model parameters, $r$ and $s$, of both systems are given in Table~\ref{tab:phi_mp}.

\begin{table}[ht!]
\caption{Values of the model parameters of Eq.~(\ref{eqn:phi}) by fitting to the data of both systems.}
\label{tab:phi_mp}
\begin{ruledtabular}
\begin{tabular}{lccc}
\textrm{System}&
\textrm{$\phi_m$}&
\textrm{$r$}&
\textrm{$s$ } \\
\colrule
Inclined surface flow & 0.64  & 0.077 & 0.163 \\
Bin flow ($I\leq 0.03$) & 0.644  & 0.256 & 0.481 \\
Bin flow ($I>0.03$) & 0.662  & 0.112 & 0.169
\end{tabular}
\end{ruledtabular}
\end{table}

\subsection{Normal stress differences}
\begin{figure}[ht!]
\centering
\includegraphics[scale=0.7]{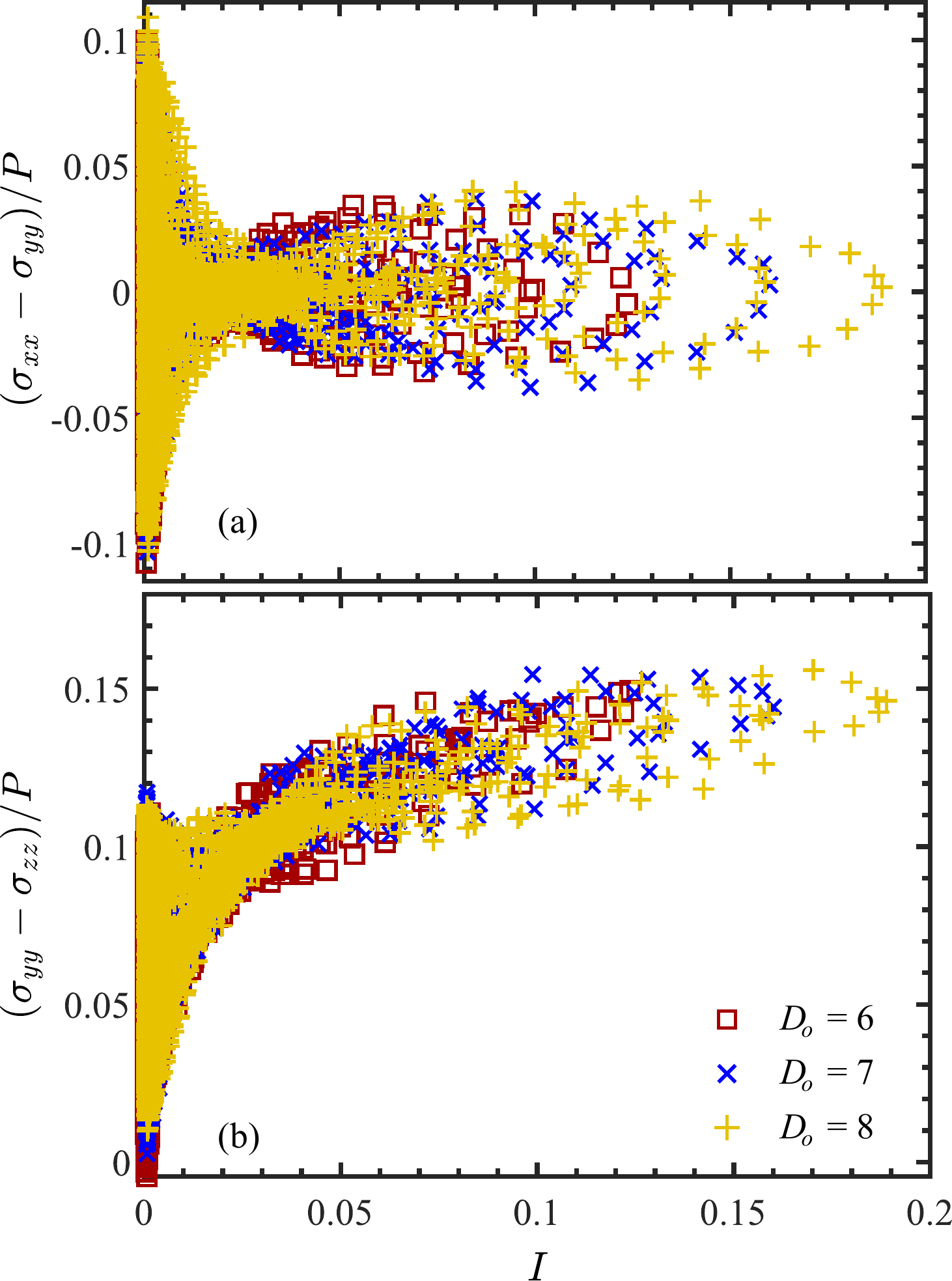}
\caption{Variation of the scaled normal stress difference (a) $N_1=(\sigma_{xx}-\sigma_{yy})/P$ and (b) $N_2=(\sigma_{yy}-\sigma_{zz})/P$  with inertial number $I$ for all outlet widths. No bound in standard error is considered while plotting the data in (a) and (b). Legend for both plots is given in (b).}
\label{fig:nsd}
\end{figure}

In the transformed coordinates, the normal stresses correspond to the components of the pressure, as shown in Eq.~(\ref{eqn:sigma}). Fig.~\ref{fig:nsd} shows the variation of the normal stress differences normalized by the pressure ($N_1=(\sigma_{xx}-\sigma_{yy})/P$ and  $N_2=(\sigma_{yy}-\sigma_{zz})/P$) with the inertial number ($I$) for bin flow for different outlet sizes. The in-plane, first normal stress difference ($N_1$, Fig.~\ref{fig:nsd}(a)) is close to zero over the range of inertial numbers studied and the second normal stress difference ($N_2$,  Fig.~\ref{fig:nsd}(b)) is positive and increases with inertial number. These results are in agreement with earlier computational \cite{ tripathi2011,campbell1986,campbell1989,alam2005} and theoretical \cite{saha2016} results for shear flow. 

\begin{figure}[ht!]
\centering
\includegraphics[scale=0.7]{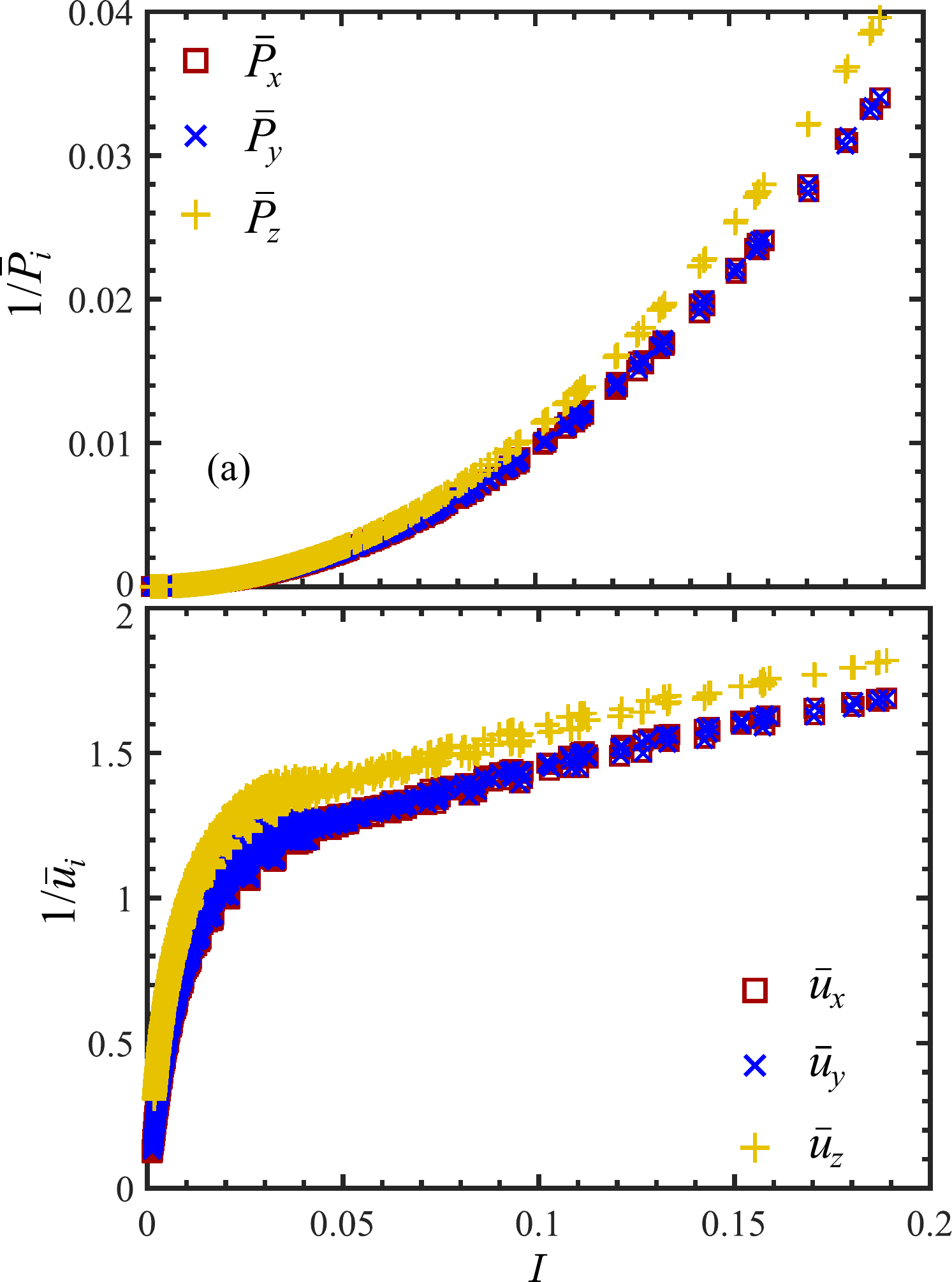}
\caption{(a) Inverse of the scaled pressure components $1/\overline{P}_i$, where $i=\{x,y,z\}$, with inertial number $I$. (b) Reciprocal of the scaled fluctuation velocity components $1/\overline{u}_i$ with $I$. The data are shown for $D_o=8$. Qualitatively similar variation is observed for other outlet widths.}
\label{fig:pfxyz}
\end{figure}

\begin{figure}[ht!]
\centering
\includegraphics[scale=0.7]{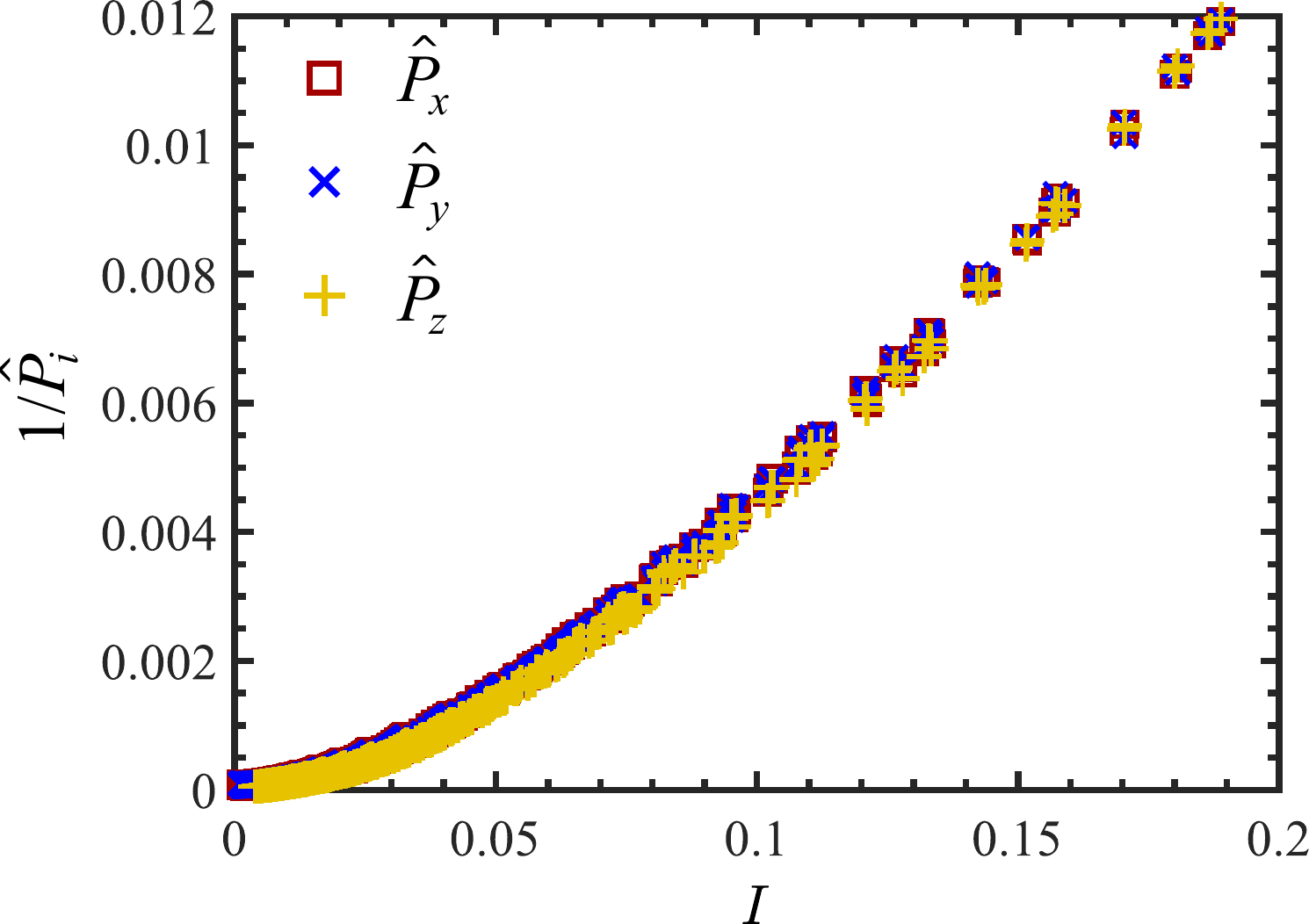}
\caption{Variation of the inverse of the scaled pressure components $1/\hat{P}_i$ with $I$. The data are shown for $D_o=8$. Qualitatively similar plots are obtained for other $D_o$ values.}
\label{fig:pxyz}
\end{figure}

Fig.~\ref{fig:pfxyz}(a) shows the variation of the scaled pressure components, $\overline{P}_{i}=P_{i}/\rho \dot{\gamma}^2 d^2$ for $i=\{x,y,z\}$, with inertial number $I$. The in-plane pressure components $\overline{P}_x$ and $\overline{P}_y$ are nearly equal, whereas the out-of-plane component $\overline{P}_z$ is smaller in comparison.  Fig.~\ref{fig:pfxyz}(b) shows the variation of the scaled fluctuation velocity components, $\overline{u}_i=u_i/\dot{\gamma}d$ for $i=\{x,y,z\}$, with inertial number ($I$). The in plane components of the fluctuation velocity ($\overline{u}_x, \overline{u}_y$) are equal and larger than the $z$-component ($\overline{u}_z$).  This observation correlates well with results for the scaled pressure components. Finally, the variation of pressure components scaled with the respective fluctuation velocities, $\hat{P}_{i}=P_{i}/\rho u_i^2$, with the inertial number ($I$) is shown in Fig.~\ref{fig:pxyz}. The data for all three components collapse to a single curve, indicating that the anisotropy of pressure is directly related to the anisotropy in the fluctuation velocity, in agreement with the theory of \citet{saha2016}.

The results presented here indicate that the in plane components of the pressure and velocity fluctuations are isotropic. Further, the pressure components scaled with the fluctuation velocity component ($\hat{P}_i$) are the same for all three directions. 

%

%

%

\section{Conclusions}
\label{sec:conclusion}
A detailed analysis of granular rheology in a steady flow from a bin, a flow with two non-zero velocity components, was presented based on DEM simulations. Three different outlet sizes were used, which gave a nearly two fold variation of the flow rate from the bin. The flow was in the slow, dense regime ($I<0.15$), with viscous effects dominant ($Re<1$). A transformation of the coordinates enabled direct computation of the friction coefficient ($\mu$), viscosity ($\eta$) and components of the pressure ($P_x, P_y, P_z$). Scaling relations were presented to enable computation of the rheological parameters, with the purpose of utilization in computational fluid dynamics (CFD) calculations of granular flows in complex geometries. Only data with a sufficiently small standard error were considered in the analysis, to ensure reliability of the results.

The data for the friction coefficient $\mu$ versus the inertial number $I$ showed a reasonably good collapse and was described by  Eq.~(\ref{eqn:jop})\cite{jop2006} for $I> 0.02$, but $\mu$ decreased sharply with $I$ at smaller values of the inertial number. In particular, values of $\mu$ much smaller than $\mu_s$ were obtained at low $I$ indicating a complex yielding behaviour. The best correlation for the viscosity ($\eta$) was obtained when it was scaled with the shear rate and plotted against the inertial number, i.e., $\hat{\eta} =\eta/\rho\dot{\gamma}d^2$ versus $I$. The collapse of the data was also the best in this case and a power law variation was obtained for $I > 0.003$. The results indicate that a power law relation, such as Eq.~(\ref{eqn:eta_I_1}), may be used over the entire range of $I$ instead of considering a yield stress for the material. A correlation for the fluctuation velocity was obtained, obviating the need for solving the energy balance equation. The solid fraction was shown to correlate reasonably with the inertial member and was well described by piecewise functions for $I>0.03$ and $I\leq0.03$.

Results presented for flow on a rough inclined surface are similar to the results for bin flow, but deviate from them. The cause of the deviation is layering of particles in the inclined surface flow, which results in a reduction in friction coefficient ($\mu$) and viscosity ($\eta$) and an increase in the volume fraction ($\phi$). Although unidirectional shear flows, such as the inclined surface flow, are convenient from the viewpoint of analysis, layering of particles and the possibility of crystallization within layers can introduce significant effects\cite{kumaran2012,mandal2017}. Using flows which do not cause such layering effects is thus recommended in studies of granular rheology.

The computed normal stress differences for the bin flow are in agreement with previous results for shear flow, with the in plane normal stress difference close to zero and the second normal stress difference positive. Thus, the pressure in the plane of the flow ($x$-$y$) is nearly isotropic, while component of the pressure in $z$-direction is slightly smaller. The pressure components correlate well with the components of the fluctuation velocity, implying that the pressure differences are due to differences in the magnitude of the fluctuation velocity components.

The results presented above indicate that the governing equations for the two-dimensional flow considered are given by Eqs.~(\ref{eqn:be1}) and (\ref{eqn:be2}), with the stress constitutive equation 
\begin{equation}
\bm{\sigma}=P\bm{I}-2\eta\bm{D}
\end{equation}
where $\bm{D}=(\bm{\nabla v}+\bm{\nabla v}^T)/2$ and $P=(\sigma_{xx}+\sigma_{yy})/2$, since the pressure is isotropic in the $x$-$y$ plane. The viscosity, $\eta$, is given by Eq.~(\ref{eqn:eta_I_1}) and the solid fraction, $\phi$, by Eq.~(\ref{eqn:phi}). The parameters in the correlations will depend on the type of particles being considered, and need to be obtained empirically, by experiments or DEM simulations as done in the present work. Thus, a simple continuum model in which the granular fluid is a slightly compressible power law fluid, describes the flow for the range of parameters studied. 


\section*{Acknowledgements}
Financial support of IIT Bombay and SERB, India (Grant No. SR/S2/JCB-34/2010) is gratefully acknowledged. We also thank Prof. N. Kumbhakarna for providing access to his cluster for performing computations. A.B. gratefully acknowledges the hospitality of the Department of Chemical Engineering, IIT Bombay on his visit in summer 2019.
\appendix
\section{Velocity profile for an inclined surface flow}
\noindent From Eq.~(\ref{eqn:eta_I_1}),
\begin{equation}
\dfrac{1}{\hat{\eta}}=mI^n,
\end{equation}
 where $\hat{\eta} =\eta/\rho \dot{\gamma}d^2$. Thus, using $I=\dot{\gamma}d/\sqrt{P/\rho}$,
\begin{equation}
m\left(\dfrac{\dot{\gamma}d}{\sqrt{P/\rho}}\right)^n = \dfrac{\rho\dot{\gamma}d^2}{\eta}.
\label{eqn:app0}
\end{equation}
Considering a steady, fully developed flow on the inclined surface with velocity profile $v_x=v_x(y)$ and assuming a constant solid fraction $\phi$, the shear stress and pressure profiles are obtained by the conservation of linear momentum as $\tau_{xy} =- \rho g (H-y) \phi\, \text{sin}\theta $ and $P = \rho g (H-y) \phi\, \text{cos}\theta$. The effective friction coefficient for the inclined surface flow may be given by $\mu=|\tau_{xy}|/P = \text{tan}\theta$, which is a constant quantity. Considering the constitutive relation for the fluid $\tau_{xy}=-\eta\, dv_x/dy$, where $\dot{\gamma}= dv_x/dy$, and $\mu=\text{tan}\theta$, we obtain
\begin{equation}
\dfrac{dv_x}{dy} = \dfrac{1}{d} \, (m\text{tan}\theta)^{1/(2-n)} \, (g\phi \text{cos}\theta)^{1/2} \, (H-y)^{1/2}.
\label{eqn:app2}
\end{equation}
Integrating Eq.~(\ref{eqn:app2}), we get
\begin{equation}
v_x = \dfrac{2}{3d} \, (m\text{tan}\theta)^{1/(2-n)} \, (g\phi \text{cos}\theta)^{1/2} \, \left\lbrace H^{3/2}-(H-y)^{1/2}\right\rbrace.
\end{equation}

\bibliography{mybib}

\providecommand{\noopsort}[1]{}\providecommand{\singleletter}[1]{#1}%
\begin{thebibliography}{88}%
\makeatletter
\providecommand \@ifxundefined [1]{%
 \@ifx{#1\undefined}
}%
\providecommand \@ifnum [1]{%
 \ifnum #1\expandafter \@firstoftwo
 \else \expandafter \@secondoftwo
 \fi
}%
\providecommand \@ifx [1]{%
 \ifx #1\expandafter \@firstoftwo
 \else \expandafter \@secondoftwo
 \fi
}%
\providecommand \natexlab [1]{#1}%
\providecommand \enquote  [1]{``#1''}%
\providecommand \bibnamefont  [1]{#1}%
\providecommand \bibfnamefont [1]{#1}%
\providecommand \citenamefont [1]{#1}%
\providecommand \href@noop [0]{\@secondoftwo}%
\providecommand \href [0]{\begingroup \@sanitize@url \@href}%
\providecommand \@href[1]{\@@startlink{#1}\@@href}%
\providecommand \@@href[1]{\endgroup#1\@@endlink}%
\providecommand \@sanitize@url [0]{\catcode `\\12\catcode `\$12\catcode
  `\&12\catcode `\#12\catcode `\^12\catcode `\_12\catcode `\%12\relax}%
\providecommand \@@startlink[1]{}%
\providecommand \@@endlink[0]{}%
\providecommand \url  [0]{\begingroup\@sanitize@url \@url }%
\providecommand \@url [1]{\endgroup\@href {#1}{\urlprefix }}%
\providecommand \urlprefix  [0]{URL }%
\providecommand \Eprint [0]{\href }%
\providecommand \doibase [0]{http://dx.doi.org/}%
\providecommand \selectlanguage [0]{\@gobble}%
\providecommand \bibinfo  [0]{\@secondoftwo}%
\providecommand \bibfield  [0]{\@secondoftwo}%
\providecommand \translation [1]{[#1]}%
\providecommand \BibitemOpen [0]{}%
\providecommand \bibitemStop [0]{}%
\providecommand \bibitemNoStop [0]{.\EOS\space}%
\providecommand \EOS [0]{\spacefactor3000\relax}%
\providecommand \BibitemShut  [1]{\csname bibitem#1\endcsname}%
\let\auto@bib@innerbib\@empty
\bibitem [{\citenamefont {Jaeger}, \citenamefont {Nagel},\ and\ \citenamefont
  {Behringer}(1996)}]{jaeger1996}%
  \BibitemOpen
  \bibfield  {author} {\bibinfo {author} {\bibfnamefont {H.~M.}\ \bibnamefont
  {Jaeger}}, \bibinfo {author} {\bibfnamefont {S.~R.}\ \bibnamefont {Nagel}}, \
  and\ \bibinfo {author} {\bibfnamefont {R.~P.}\ \bibnamefont {Behringer}},\
  }\bibfield  {title} {\enquote {\bibinfo {title} {Granular solids, liquids,
  and gases},}\ }\href@noop {} {\bibfield  {journal} {\bibinfo  {journal} {Rev.
  Mod. Phys.}\ }\textbf {\bibinfo {volume} {68}},\ \bibinfo {pages} {1259}
  (\bibinfo {year} {1996})}\BibitemShut {NoStop}%
\bibitem [{\citenamefont {Andreotti}, \citenamefont {Forterre},\ and\
  \citenamefont {Pouliquen}(2013)}]{pouliquen_book}%
  \BibitemOpen
  \bibfield  {author} {\bibinfo {author} {\bibfnamefont {B.}~\bibnamefont
  {Andreotti}}, \bibinfo {author} {\bibfnamefont {Y.}~\bibnamefont {Forterre}},
  \ and\ \bibinfo {author} {\bibfnamefont {O.}~\bibnamefont {Pouliquen}},\
  }\href@noop {} {\emph {\bibinfo {title} {Granular Media: {B}etween Fluid and
  Solid}}}\ (\bibinfo  {publisher} {Cambridge University Press},\ \bibinfo
  {year} {2013})\BibitemShut {NoStop}%
\bibitem [{\citenamefont {Ottino}\ and\ \citenamefont
  {Khakhar}(2000)}]{ottino2000}%
  \BibitemOpen
  \bibfield  {author} {\bibinfo {author} {\bibfnamefont {J.~M.}\ \bibnamefont
  {Ottino}}\ and\ \bibinfo {author} {\bibfnamefont {D.~V.}\ \bibnamefont
  {Khakhar}},\ }\bibfield  {title} {\enquote {\bibinfo {title} {Mixing and
  segregation of granular materials},}\ }\href@noop {} {\bibfield  {journal}
  {\bibinfo  {journal} {Ann. Rev. Fluid Mech.}\ }\textbf {\bibinfo {volume}
  {32}},\ \bibinfo {pages} {55--91} (\bibinfo {year} {2000})}\BibitemShut
  {NoStop}%
\bibitem [{\citenamefont {Cordero}\ and\ \citenamefont
  {Pugnaloni}(2015)}]{cordero2015}%
  \BibitemOpen
  \bibfield  {author} {\bibinfo {author} {\bibfnamefont {M.~J.}\ \bibnamefont
  {Cordero}}\ and\ \bibinfo {author} {\bibfnamefont {L.~A.}\ \bibnamefont
  {Pugnaloni}},\ }\bibfield  {title} {\enquote {\bibinfo {title} {Dynamic
  transition in conveyor belt driven granular flow},}\ }\href@noop {}
  {\bibfield  {journal} {\bibinfo  {journal} {Powder technology}\ }\textbf
  {\bibinfo {volume} {272}},\ \bibinfo {pages} {290--294} (\bibinfo {year}
  {2015})}\BibitemShut {NoStop}%
\bibitem [{\citenamefont {Zhu}\ \emph {et~al.}(2019)\citenamefont {Zhu},
  \citenamefont {Wang}, \citenamefont {Shi}, \citenamefont {Li},\ and\
  \citenamefont {Zheng}}]{zhu2019}%
  \BibitemOpen
  \bibfield  {author} {\bibinfo {author} {\bibfnamefont {H.-W.}\ \bibnamefont
  {Zhu}}, \bibinfo {author} {\bibfnamefont {L.-P.}\ \bibnamefont {Wang}},
  \bibinfo {author} {\bibfnamefont {Q.-F.}\ \bibnamefont {Shi}}, \bibinfo
  {author} {\bibfnamefont {L.-S.}\ \bibnamefont {Li}}, \ and\ \bibinfo {author}
  {\bibfnamefont {N.}~\bibnamefont {Zheng}},\ }\bibfield  {title} {\enquote
  {\bibinfo {title} {Improvement in flow rate through an aperture on a conveyor
  belt: Effects of bottom wall and packing configurations},}\ }\href@noop {}
  {\bibfield  {journal} {\bibinfo  {journal} {Powder Technology}\ }\textbf
  {\bibinfo {volume} {345}},\ \bibinfo {pages} {676--681} (\bibinfo {year}
  {2019})}\BibitemShut {NoStop}%
\bibitem [{\citenamefont {Janda}, \citenamefont {Zuriguel},\ and\ \citenamefont
  {Maza}(2012)}]{janda2012}%
  \BibitemOpen
  \bibfield  {author} {\bibinfo {author} {\bibfnamefont {A.}~\bibnamefont
  {Janda}}, \bibinfo {author} {\bibfnamefont {I.}~\bibnamefont {Zuriguel}}, \
  and\ \bibinfo {author} {\bibfnamefont {D.}~\bibnamefont {Maza}},\ }\bibfield
  {title} {\enquote {\bibinfo {title} {Flow rate of particles through apertures
  obtained from self-similar density and velocity profiles},}\ }\href@noop {}
  {\bibfield  {journal} {\bibinfo  {journal} {Phys. Rev. Lett.}\ }\textbf
  {\bibinfo {volume} {108}},\ \bibinfo {pages} {248001} (\bibinfo {year}
  {2012})}\BibitemShut {NoStop}%
\bibitem [{\citenamefont {Peralta}\ \emph {et~al.}(2017)\citenamefont
  {Peralta}, \citenamefont {Aguirre}, \citenamefont {G{\'e}minard},\ and\
  \citenamefont {Pugnaloni}}]{peralta2017}%
  \BibitemOpen
  \bibfield  {author} {\bibinfo {author} {\bibfnamefont {J.~P.}\ \bibnamefont
  {Peralta}}, \bibinfo {author} {\bibfnamefont {M.~A.}\ \bibnamefont
  {Aguirre}}, \bibinfo {author} {\bibfnamefont {J.-C.}\ \bibnamefont
  {G{\'e}minard}}, \ and\ \bibinfo {author} {\bibfnamefont {L.~A.}\
  \bibnamefont {Pugnaloni}},\ }\bibfield  {title} {\enquote {\bibinfo {title}
  {Apparent mass during silo discharge: {N}onlinear effects related to filling
  protocols},}\ }\href@noop {} {\bibfield  {journal} {\bibinfo  {journal}
  {Powder Technology}\ }\textbf {\bibinfo {volume} {311}},\ \bibinfo {pages}
  {265--272} (\bibinfo {year} {2017})}\BibitemShut {NoStop}%
\bibitem [{\citenamefont {Takahashi}(1981)}]{takahashi1981}%
  \BibitemOpen
  \bibfield  {author} {\bibinfo {author} {\bibfnamefont {T.}~\bibnamefont
  {Takahashi}},\ }\bibfield  {title} {\enquote {\bibinfo {title} {Debris
  flow},}\ }\href@noop {} {\bibfield  {journal} {\bibinfo  {journal} {Annu.
  Rev. Fluid Mech.}\ }\textbf {\bibinfo {volume} {13}},\ \bibinfo {pages}
  {57--77} (\bibinfo {year} {1981})}\BibitemShut {NoStop}%
\bibitem [{\citenamefont {Ancey}(2007)}]{ancey2007}%
  \BibitemOpen
  \bibfield  {author} {\bibinfo {author} {\bibfnamefont {C.}~\bibnamefont
  {Ancey}},\ }\bibfield  {title} {\enquote {\bibinfo {title} {Plasticity and
  geophysical flows: {A} review},}\ }\href@noop {} {\bibfield  {journal}
  {\bibinfo  {journal} {J. Non-Newtonian Fluid Mech.}\ }\textbf {\bibinfo
  {volume} {142}},\ \bibinfo {pages} {4--35} (\bibinfo {year}
  {2007})}\BibitemShut {NoStop}%
\bibitem [{\citenamefont {Cundall}\ and\ \citenamefont
  {Strack}(1979)}]{cundall1979}%
  \BibitemOpen
  \bibfield  {author} {\bibinfo {author} {\bibfnamefont {P.~A.}\ \bibnamefont
  {Cundall}}\ and\ \bibinfo {author} {\bibfnamefont {O.~D.~L.}\ \bibnamefont
  {Strack}},\ }\bibfield  {title} {\enquote {\bibinfo {title} {A discrete
  numerical model for granular assemblies},}\ }\href@noop {} {\bibfield
  {journal} {\bibinfo  {journal} {Geotechnique}\ }\textbf {\bibinfo {volume}
  {29(1)}},\ \bibinfo {pages} {47--65} (\bibinfo {year} {1979})}\BibitemShut
  {NoStop}%
\bibitem [{\citenamefont {P\"{o}schel}\ and\ \citenamefont
  {Schwager}(2005)}]{poschel}%
  \BibitemOpen
  \bibfield  {author} {\bibinfo {author} {\bibfnamefont {T.}~\bibnamefont
  {P\"{o}schel}}\ and\ \bibinfo {author} {\bibfnamefont {T.}~\bibnamefont
  {Schwager}},\ }\href@noop {} {\emph {\bibinfo {title} {Computational Granular
  Dynamics: Models and Algorithms}}}\ (\bibinfo  {publisher} {Springer},\
  \bibinfo {year} {2005})\BibitemShut {NoStop}%
\bibitem [{\citenamefont {Thornton}(2015)}]{thornton}%
  \BibitemOpen
  \bibfield  {author} {\bibinfo {author} {\bibfnamefont {C.}~\bibnamefont
  {Thornton}},\ }\href@noop {} {\emph {\bibinfo {title} {Granular Dynamics,
  Contact Mechanics and Particle System Simulations}}}\ (\bibinfo  {publisher}
  {Springer},\ \bibinfo {year} {2015})\BibitemShut {NoStop}%
\bibitem [{\citenamefont {Mishra}(2003)}]{bkm2003}%
  \BibitemOpen
  \bibfield  {author} {\bibinfo {author} {\bibfnamefont {B.~K.}\ \bibnamefont
  {Mishra}},\ }\bibfield  {title} {\enquote {\bibinfo {title} {A review of
  computer simulation of tumbling mills by the discrete element method: Part
  {I}-contact mechanics},}\ }\href@noop {} {\bibfield  {journal} {\bibinfo
  {journal} {Int. J. Miner. Process.}\ }\textbf {\bibinfo {volume} {71}},\
  \bibinfo {pages} {73--93} (\bibinfo {year} {2003})}\BibitemShut {NoStop}%
\bibitem [{\citenamefont {Guo}\ and\ \citenamefont {Curtis}(2015)}]{guo2015}%
  \BibitemOpen
  \bibfield  {author} {\bibinfo {author} {\bibfnamefont {Y.}~\bibnamefont
  {Guo}}\ and\ \bibinfo {author} {\bibfnamefont {J.~S.}\ \bibnamefont
  {Curtis}},\ }\bibfield  {title} {\enquote {\bibinfo {title} {Discrete element
  method simulations for complex granular flows},}\ }\href@noop {} {\bibfield
  {journal} {\bibinfo  {journal} {Ann. Rev. Fluid Mech.}\ }\textbf {\bibinfo
  {volume} {47}},\ \bibinfo {pages} {21--46} (\bibinfo {year}
  {2015})}\BibitemShut {NoStop}%
\bibitem [{\citenamefont {Bhateja}, \citenamefont {Sharma},\ and\ \citenamefont
  {Singh}(2017)}]{bhateja2017}%
  \BibitemOpen
  \bibfield  {author} {\bibinfo {author} {\bibfnamefont {A.}~\bibnamefont
  {Bhateja}}, \bibinfo {author} {\bibfnamefont {I.}~\bibnamefont {Sharma}}, \
  and\ \bibinfo {author} {\bibfnamefont {J.~K.}\ \bibnamefont {Singh}},\
  }\bibfield  {title} {\enquote {\bibinfo {title} {Segregation physics of a
  macroscale granular ratchet},}\ }\href@noop {} {\bibfield  {journal}
  {\bibinfo  {journal} {Phys. Rev. Fluids}\ }\textbf {\bibinfo {volume} {2}},\
  \bibinfo {pages} {052301} (\bibinfo {year} {2017})}\BibitemShut {NoStop}%
\bibitem [{\citenamefont {Cleary}\ and\ \citenamefont
  {Sawley}(2002)}]{cleary2002}%
  \BibitemOpen
  \bibfield  {author} {\bibinfo {author} {\bibfnamefont {P.~W.}\ \bibnamefont
  {Cleary}}\ and\ \bibinfo {author} {\bibfnamefont {M.~L.}\ \bibnamefont
  {Sawley}},\ }\bibfield  {title} {\enquote {\bibinfo {title} {D{EM} modelling
  of industrial granular flows: {3D} case studies and the effect of particle
  shape on hopper discharge},}\ }\href@noop {} {\bibfield  {journal} {\bibinfo
  {journal} {Appl. Math. Modell.}\ }\textbf {\bibinfo {volume} {26}},\ \bibinfo
  {pages} {89--111} (\bibinfo {year} {2002})}\BibitemShut {NoStop}%
\bibitem [{\citenamefont {Cleary}(2004)}]{cleary2004}%
  \BibitemOpen
  \bibfield  {author} {\bibinfo {author} {\bibfnamefont {P.~W.}\ \bibnamefont
  {Cleary}},\ }\bibfield  {title} {\enquote {\bibinfo {title} {Large scale
  industrial {DEM} modelling},}\ }\href@noop {} {\bibfield  {journal} {\bibinfo
   {journal} {Eng. Comput.}\ }\textbf {\bibinfo {volume} {21}},\ \bibinfo
  {pages} {169--204} (\bibinfo {year} {2004})}\BibitemShut {NoStop}%
\bibitem [{\citenamefont {Cleary}(2010)}]{cleary2010}%
  \BibitemOpen
  \bibfield  {author} {\bibinfo {author} {\bibfnamefont {P.~W.}\ \bibnamefont
  {Cleary}},\ }\bibfield  {title} {\enquote {\bibinfo {title} {{DEM} prediction
  of industrial and geophysical particle flows},}\ }\href@noop {} {\bibfield
  {journal} {\bibinfo  {journal} {Particuology}\ }\textbf {\bibinfo {volume}
  {8}},\ \bibinfo {pages} {106--118} (\bibinfo {year} {2010})}\BibitemShut
  {NoStop}%
\bibitem [{\citenamefont {Forterre}\ and\ \citenamefont
  {Pouliquen}(2018)}]{forterre2018}%
  \BibitemOpen
  \bibfield  {author} {\bibinfo {author} {\bibfnamefont {Y.}~\bibnamefont
  {Forterre}}\ and\ \bibinfo {author} {\bibfnamefont {O.}~\bibnamefont
  {Pouliquen}},\ }\bibfield  {title} {\enquote {\bibinfo {title} {Physics of
  particulate flows: From sand avalanche to active suspensions in plants},}\
  }\href@noop {} {\bibfield  {journal} {\bibinfo  {journal} {C. R. Physique}\
  }\textbf {\bibinfo {volume} {19}},\ \bibinfo {pages} {271--284} (\bibinfo
  {year} {2018})}\BibitemShut {NoStop}%
\bibitem [{\citenamefont {Forterre}\ and\ \citenamefont
  {Pouliquen}(2008)}]{forterre2008}%
  \BibitemOpen
  \bibfield  {author} {\bibinfo {author} {\bibfnamefont {Y.}~\bibnamefont
  {Forterre}}\ and\ \bibinfo {author} {\bibfnamefont {O.}~\bibnamefont
  {Pouliquen}},\ }\bibfield  {title} {\enquote {\bibinfo {title} {Flows of
  dense granular media},}\ }\href@noop {} {\bibfield  {journal} {\bibinfo
  {journal} {Annu. Rev. Fluid Mech.}\ }\textbf {\bibinfo {volume} {40}},\
  \bibinfo {pages} {1--24} (\bibinfo {year} {2008})}\BibitemShut {NoStop}%
\bibitem [{\citenamefont {Larson}(1999)}]{larson}%
  \BibitemOpen
  \bibfield  {author} {\bibinfo {author} {\bibfnamefont {R.~G.}\ \bibnamefont
  {Larson}},\ }\href@noop {} {\emph {\bibinfo {title} {The structure and
  rheology of complex fluids}}}\ (\bibinfo  {publisher} {Oxford University
  Press},\ \bibinfo {year} {1999})\BibitemShut {NoStop}%
\bibitem [{\citenamefont {Bagnold}(1954)}]{bagnold1954}%
  \BibitemOpen
  \bibfield  {author} {\bibinfo {author} {\bibfnamefont {R.~A.}\ \bibnamefont
  {Bagnold}},\ }\bibfield  {title} {\enquote {\bibinfo {title} {Experiments on
  a gravity-free dispersion of large solid spheres in a newtonian fluid under
  shear},}\ }\href@noop {} {\bibfield  {journal} {\bibinfo  {journal} {Proc. R.
  Soc. Lond. A}\ }\textbf {\bibinfo {volume} {225}},\ \bibinfo {pages} {49--63}
  (\bibinfo {year} {1954})}\BibitemShut {NoStop}%
\bibitem [{\citenamefont {Haff}(1983)}]{haff1983}%
  \BibitemOpen
  \bibfield  {author} {\bibinfo {author} {\bibfnamefont {P.~K.}\ \bibnamefont
  {Haff}},\ }\bibfield  {title} {\enquote {\bibinfo {title} {Grain flow as a
  fluid-mechanical phenomenon},}\ }\href@noop {} {\bibfield  {journal}
  {\bibinfo  {journal} {J. Fluid Mech.}\ }\textbf {\bibinfo {volume} {134}},\
  \bibinfo {pages} {401--430} (\bibinfo {year} {1983})}\BibitemShut {NoStop}%
\bibitem [{\citenamefont {Jenkins}\ and\ \citenamefont
  {Savage}(1983)}]{jenkins1983}%
  \BibitemOpen
  \bibfield  {author} {\bibinfo {author} {\bibfnamefont {J.~T.}\ \bibnamefont
  {Jenkins}}\ and\ \bibinfo {author} {\bibfnamefont {S.~B.}\ \bibnamefont
  {Savage}},\ }\bibfield  {title} {\enquote {\bibinfo {title} {A theory for the
  rapid flow of identical, smooth, nearly elastic, spherical particles},}\
  }\href@noop {} {\bibfield  {journal} {\bibinfo  {journal} {J. Fluid Mech.}\
  }\textbf {\bibinfo {volume} {130}},\ \bibinfo {pages} {187--202} (\bibinfo
  {year} {1983})}\BibitemShut {NoStop}%
\bibitem [{\citenamefont {Lun}\ \emph {et~al.}(1984)\citenamefont {Lun},
  \citenamefont {Savage}, \citenamefont {Jeffrey},\ and\ \citenamefont
  {Chepurniy}}]{lun1984}%
  \BibitemOpen
  \bibfield  {author} {\bibinfo {author} {\bibfnamefont {C.~K.~K.}\
  \bibnamefont {Lun}}, \bibinfo {author} {\bibfnamefont {S.~B.}\ \bibnamefont
  {Savage}}, \bibinfo {author} {\bibfnamefont {D.~J.}\ \bibnamefont {Jeffrey}},
  \ and\ \bibinfo {author} {\bibfnamefont {N.}~\bibnamefont {Chepurniy}},\
  }\bibfield  {title} {\enquote {\bibinfo {title} {Kinetic theories for
  granular flow: inelastic particles in couette flow and slightly inelastic
  particles in a general flowfield},}\ }\href@noop {} {\bibfield  {journal}
  {\bibinfo  {journal} {J. Fluid Mech.}\ }\textbf {\bibinfo {volume} {140}},\
  \bibinfo {pages} {223--256} (\bibinfo {year} {1984})}\BibitemShut {NoStop}%
\bibitem [{\citenamefont {Pouliquen}(1999)}]{pouliquen1999}%
  \BibitemOpen
  \bibfield  {author} {\bibinfo {author} {\bibfnamefont {O.}~\bibnamefont
  {Pouliquen}},\ }\bibfield  {title} {\enquote {\bibinfo {title} {Scaling laws
  in granular flows down rough inclined planes},}\ }\href@noop {} {\bibfield
  {journal} {\bibinfo  {journal} {Phys. fluids}\ }\textbf {\bibinfo {volume}
  {11}},\ \bibinfo {pages} {542--548} (\bibinfo {year} {1999})}\BibitemShut
  {NoStop}%
\bibitem [{\citenamefont {Silbert}\ \emph {et~al.}(2001)\citenamefont
  {Silbert}, \citenamefont {Erta{\c{s}}}, \citenamefont {Grest}, \citenamefont
  {Halsey}, \citenamefont {Levine},\ and\ \citenamefont
  {Plimpton}}]{silbert2001}%
  \BibitemOpen
  \bibfield  {author} {\bibinfo {author} {\bibfnamefont {L.~E.}\ \bibnamefont
  {Silbert}}, \bibinfo {author} {\bibfnamefont {D.}~\bibnamefont
  {Erta{\c{s}}}}, \bibinfo {author} {\bibfnamefont {G.~S.}\ \bibnamefont
  {Grest}}, \bibinfo {author} {\bibfnamefont {T.~C.}\ \bibnamefont {Halsey}},
  \bibinfo {author} {\bibfnamefont {D.}~\bibnamefont {Levine}}, \ and\ \bibinfo
  {author} {\bibfnamefont {S.~J.}\ \bibnamefont {Plimpton}},\ }\bibfield
  {title} {\enquote {\bibinfo {title} {Granular flow down an inclined plane:
  Bagnold scaling and rheology},}\ }\href@noop {} {\bibfield  {journal}
  {\bibinfo  {journal} {Phys. Rev. E}\ }\textbf {\bibinfo {volume} {64}},\
  \bibinfo {pages} {051302} (\bibinfo {year} {2001})}\BibitemShut {NoStop}%
\bibitem [{\citenamefont {MiDi}(2004)}]{midi2004}%
  \BibitemOpen
  \bibfield  {author} {\bibinfo {author} {\bibfnamefont {{\relax
  GDR}.}~\bibnamefont {MiDi}},\ }\bibfield  {title} {\enquote {\bibinfo {title}
  {On dense granular flows},}\ }\href@noop {} {\bibfield  {journal} {\bibinfo
  {journal} {Eur. Phys. J. E}\ }\textbf {\bibinfo {volume} {14}},\ \bibinfo
  {pages} {341--365} (\bibinfo {year} {2004})}\BibitemShut {NoStop}%
\bibitem [{\citenamefont {Pouliquen}(2004)}]{pouliquen2004}%
  \BibitemOpen
  \bibfield  {author} {\bibinfo {author} {\bibfnamefont {O.}~\bibnamefont
  {Pouliquen}},\ }\bibfield  {title} {\enquote {\bibinfo {title} {Velocity
  correlations in dense granular flows},}\ }\href@noop {} {\bibfield  {journal}
  {\bibinfo  {journal} {Phys. Rev. Lett.}\ }\textbf {\bibinfo {volume} {93}},\
  \bibinfo {pages} {248001} (\bibinfo {year} {2004})}\BibitemShut {NoStop}%
\bibitem [{\citenamefont {da~Cruz}\ \emph {et~al.}(2005)\citenamefont
  {da~Cruz}, \citenamefont {Emam}, \citenamefont {Prochnow}, \citenamefont
  {Roux},\ and\ \citenamefont {Chevoir}}]{dacruz2005}%
  \BibitemOpen
  \bibfield  {author} {\bibinfo {author} {\bibfnamefont {F.}~\bibnamefont
  {da~Cruz}}, \bibinfo {author} {\bibfnamefont {S.}~\bibnamefont {Emam}},
  \bibinfo {author} {\bibfnamefont {M.}~\bibnamefont {Prochnow}}, \bibinfo
  {author} {\bibfnamefont {J.-N.}\ \bibnamefont {Roux}}, \ and\ \bibinfo
  {author} {\bibfnamefont {F.}~\bibnamefont {Chevoir}},\ }\bibfield  {title}
  {\enquote {\bibinfo {title} {Rheophysics of dense granular materials:
  Discrete simulation of plane shear flows},}\ }\href@noop {} {\bibfield
  {journal} {\bibinfo  {journal} {Phys. Rev. E}\ }\textbf {\bibinfo {volume}
  {72}},\ \bibinfo {pages} {021309} (\bibinfo {year} {2005})}\BibitemShut
  {NoStop}%
\bibitem [{\citenamefont {Jop}, \citenamefont {Forterre},\ and\ \citenamefont
  {Pouliquen}(2006)}]{jop2006}%
  \BibitemOpen
  \bibfield  {author} {\bibinfo {author} {\bibfnamefont {P.}~\bibnamefont
  {Jop}}, \bibinfo {author} {\bibfnamefont {Y.}~\bibnamefont {Forterre}}, \
  and\ \bibinfo {author} {\bibfnamefont {O.}~\bibnamefont {Pouliquen}},\
  }\bibfield  {title} {\enquote {\bibinfo {title} {A constitutive law for dense
  granular flows},}\ }\href@noop {} {\bibfield  {journal} {\bibinfo  {journal}
  {Nature}\ }\textbf {\bibinfo {volume} {441}},\ \bibinfo {pages} {727--730}
  (\bibinfo {year} {2006})}\BibitemShut {NoStop}%
\bibitem [{\citenamefont {Pouliquen}\ and\ \citenamefont
  {Forterre}(2009)}]{pouliquen2009}%
  \BibitemOpen
  \bibfield  {author} {\bibinfo {author} {\bibfnamefont {O.}~\bibnamefont
  {Pouliquen}}\ and\ \bibinfo {author} {\bibfnamefont {Y.}~\bibnamefont
  {Forterre}},\ }\bibfield  {title} {\enquote {\bibinfo {title} {A non-local
  rheology for dense granular flows},}\ }\href@noop {} {\bibfield  {journal}
  {\bibinfo  {journal} {Phil. Trans. R. Soc. A}\ }\textbf {\bibinfo {volume}
  {367}},\ \bibinfo {pages} {5091--5107} (\bibinfo {year} {2009})}\BibitemShut
  {NoStop}%
\bibitem [{\citenamefont {Staron}\ \emph {et~al.}(2010)\citenamefont {Staron},
  \citenamefont {Lagr{\'e}e}, \citenamefont {Josserand},\ and\ \citenamefont
  {Lhuillier}}]{staron2010}%
  \BibitemOpen
  \bibfield  {author} {\bibinfo {author} {\bibfnamefont {L.}~\bibnamefont
  {Staron}}, \bibinfo {author} {\bibfnamefont {P.-Y.}\ \bibnamefont
  {Lagr{\'e}e}}, \bibinfo {author} {\bibfnamefont {C.}~\bibnamefont
  {Josserand}}, \ and\ \bibinfo {author} {\bibfnamefont {D.}~\bibnamefont
  {Lhuillier}},\ }\bibfield  {title} {\enquote {\bibinfo {title} {Flow and
  jamming of a two-dimensional granular bed: Toward a nonlocal rheology?}}\
  }\href@noop {} {\bibfield  {journal} {\bibinfo  {journal} {Phys. Fluids}\
  }\textbf {\bibinfo {volume} {22}},\ \bibinfo {pages} {113303} (\bibinfo
  {year} {2010})}\BibitemShut {NoStop}%
\bibitem [{\citenamefont {Tripathi}\ and\ \citenamefont
  {Khakhar}(2011)}]{tripathi2011}%
  \BibitemOpen
  \bibfield  {author} {\bibinfo {author} {\bibfnamefont {A.}~\bibnamefont
  {Tripathi}}\ and\ \bibinfo {author} {\bibfnamefont {D.~V.}\ \bibnamefont
  {Khakhar}},\ }\bibfield  {title} {\enquote {\bibinfo {title} {Rheology of
  binary granular mixtures in the dense flow regime},}\ }\href@noop {}
  {\bibfield  {journal} {\bibinfo  {journal} {Phys. Fluids}\ }\textbf {\bibinfo
  {volume} {23}},\ \bibinfo {pages} {113302} (\bibinfo {year}
  {2011})}\BibitemShut {NoStop}%
\bibitem [{\citenamefont {Kamrin}\ and\ \citenamefont
  {Koval}(2012)}]{kamrin2012}%
  \BibitemOpen
  \bibfield  {author} {\bibinfo {author} {\bibfnamefont {K.}~\bibnamefont
  {Kamrin}}\ and\ \bibinfo {author} {\bibfnamefont {G.}~\bibnamefont {Koval}},\
  }\bibfield  {title} {\enquote {\bibinfo {title} {Nonlocal constitutive
  relation for steady granular flow},}\ }\href@noop {} {\bibfield  {journal}
  {\bibinfo  {journal} {Phys. Rev. Lett.}\ }\textbf {\bibinfo {volume} {108}},\
  \bibinfo {pages} {178301} (\bibinfo {year} {2012})}\BibitemShut {NoStop}%
\bibitem [{\citenamefont {Bouzid}\ \emph {et~al.}(2013)\citenamefont {Bouzid},
  \citenamefont {Trulsson}, \citenamefont {Claudin}, \citenamefont
  {Cl{\'e}ment},\ and\ \citenamefont {Andreotti}}]{bouzid2013}%
  \BibitemOpen
  \bibfield  {author} {\bibinfo {author} {\bibfnamefont {M.}~\bibnamefont
  {Bouzid}}, \bibinfo {author} {\bibfnamefont {M.}~\bibnamefont {Trulsson}},
  \bibinfo {author} {\bibfnamefont {P.}~\bibnamefont {Claudin}}, \bibinfo
  {author} {\bibfnamefont {E.}~\bibnamefont {Cl{\'e}ment}}, \ and\ \bibinfo
  {author} {\bibfnamefont {B.}~\bibnamefont {Andreotti}},\ }\bibfield  {title}
  {\enquote {\bibinfo {title} {Nonlocal rheology of granular flows across yield
  conditions},}\ }\href@noop {} {\bibfield  {journal} {\bibinfo  {journal}
  {Phys. Rev. Lett.}\ }\textbf {\bibinfo {volume} {111}},\ \bibinfo {pages}
  {238301} (\bibinfo {year} {2013})}\BibitemShut {NoStop}%
\bibitem [{\citenamefont {Bouzid}\ \emph {et~al.}(2015)\citenamefont {Bouzid},
  \citenamefont {Izzet}, \citenamefont {Trulsson}, \citenamefont {Cl{\'e}ment},
  \citenamefont {Claudin},\ and\ \citenamefont {Andreotti}}]{bouzid2015}%
  \BibitemOpen
  \bibfield  {author} {\bibinfo {author} {\bibfnamefont {M.}~\bibnamefont
  {Bouzid}}, \bibinfo {author} {\bibfnamefont {A.}~\bibnamefont {Izzet}},
  \bibinfo {author} {\bibfnamefont {M.}~\bibnamefont {Trulsson}}, \bibinfo
  {author} {\bibfnamefont {E.}~\bibnamefont {Cl{\'e}ment}}, \bibinfo {author}
  {\bibfnamefont {P.}~\bibnamefont {Claudin}}, \ and\ \bibinfo {author}
  {\bibfnamefont {B.}~\bibnamefont {Andreotti}},\ }\bibfield  {title} {\enquote
  {\bibinfo {title} {Non-local rheology in dense granular flows},}\ }\href@noop
  {} {\bibfield  {journal} {\bibinfo  {journal} {Eur. Phys. J. E}\ }\textbf
  {\bibinfo {volume} {38}},\ \bibinfo {pages} {1--15} (\bibinfo {year}
  {2015})}\BibitemShut {NoStop}%
\bibitem [{\citenamefont {Seguin}\ \emph {et~al.}(2016)\citenamefont {Seguin},
  \citenamefont {Coulais}, \citenamefont {Martinez}, \citenamefont {Bertho},\
  and\ \citenamefont {Gondret}}]{seguin2016}%
  \BibitemOpen
  \bibfield  {author} {\bibinfo {author} {\bibfnamefont {A.}~\bibnamefont
  {Seguin}}, \bibinfo {author} {\bibfnamefont {C.}~\bibnamefont {Coulais}},
  \bibinfo {author} {\bibfnamefont {F.}~\bibnamefont {Martinez}}, \bibinfo
  {author} {\bibfnamefont {Y.}~\bibnamefont {Bertho}}, \ and\ \bibinfo {author}
  {\bibfnamefont {P.}~\bibnamefont {Gondret}},\ }\bibfield  {title} {\enquote
  {\bibinfo {title} {Local rheological measurements in the granular flow around
  an intruder},}\ }\href@noop {} {\bibfield  {journal} {\bibinfo  {journal}
  {Phys. Rev. E}\ }\textbf {\bibinfo {volume} {93}},\ \bibinfo {pages} {012904}
  (\bibinfo {year} {2016})}\BibitemShut {NoStop}%
\bibitem [{\citenamefont {Mandal}\ and\ \citenamefont
  {Khakhar}(2016)}]{mandal2016}%
  \BibitemOpen
  \bibfield  {author} {\bibinfo {author} {\bibfnamefont {S.}~\bibnamefont
  {Mandal}}\ and\ \bibinfo {author} {\bibfnamefont {D.~V.}\ \bibnamefont
  {Khakhar}},\ }\bibfield  {title} {\enquote {\bibinfo {title} {A study of the
  rheology of planar granular flow of dumbbells using discrete element method
  simulations},}\ }\href@noop {} {\bibfield  {journal} {\bibinfo  {journal}
  {Phys. Fluids}\ }\textbf {\bibinfo {volume} {28}},\ \bibinfo {pages} {103301}
  (\bibinfo {year} {2016})}\BibitemShut {NoStop}%
\bibitem [{\citenamefont {Zhang}\ and\ \citenamefont
  {Kamrin}(2017)}]{kamrin2017}%
  \BibitemOpen
  \bibfield  {author} {\bibinfo {author} {\bibfnamefont {Q.}~\bibnamefont
  {Zhang}}\ and\ \bibinfo {author} {\bibfnamefont {K.}~\bibnamefont {Kamrin}},\
  }\bibfield  {title} {\enquote {\bibinfo {title} {Microscopic description of
  the granular fluidity field in nonlocal flow modeling},}\ }\href@noop {}
  {\bibfield  {journal} {\bibinfo  {journal} {Phys. Rev. Lett.}\ }\textbf
  {\bibinfo {volume} {118}},\ \bibinfo {pages} {058001} (\bibinfo {year}
  {2017})}\BibitemShut {NoStop}%
\bibitem [{\citenamefont {Mandal}\ and\ \citenamefont
  {Khakhar}(2017)}]{mandal2017}%
  \BibitemOpen
  \bibfield  {author} {\bibinfo {author} {\bibfnamefont {S.}~\bibnamefont
  {Mandal}}\ and\ \bibinfo {author} {\bibfnamefont {D.~V.}\ \bibnamefont
  {Khakhar}},\ }\bibfield  {title} {\enquote {\bibinfo {title}
  {Sidewall-friction-driven ordering transition in granular channel flows:
  Implications for granular rheology},}\ }\href@noop {} {\bibfield  {journal}
  {\bibinfo  {journal} {Phys. Rev. E}\ }\textbf {\bibinfo {volume} {96}},\
  \bibinfo {pages} {050901} (\bibinfo {year} {2017})}\BibitemShut {NoStop}%
\bibitem [{\citenamefont {de~Coulomb}\ \emph {et~al.}(2017)\citenamefont
  {de~Coulomb}, \citenamefont {Bouzid}, \citenamefont {Claudin}, \citenamefont
  {Cl{\'e}ment},\ and\ \citenamefont {Andreotti}}]{de2017}%
  \BibitemOpen
  \bibfield  {author} {\bibinfo {author} {\bibfnamefont {A.~F.}\ \bibnamefont
  {de~Coulomb}}, \bibinfo {author} {\bibfnamefont {M.}~\bibnamefont {Bouzid}},
  \bibinfo {author} {\bibfnamefont {P.}~\bibnamefont {Claudin}}, \bibinfo
  {author} {\bibfnamefont {E.}~\bibnamefont {Cl{\'e}ment}}, \ and\ \bibinfo
  {author} {\bibfnamefont {B.}~\bibnamefont {Andreotti}},\ }\bibfield  {title}
  {\enquote {\bibinfo {title} {Rheology of granular flows across the transition
  from soft to rigid particles},}\ }\href@noop {} {\bibfield  {journal}
  {\bibinfo  {journal} {Phys. Rev. Fluids}\ }\textbf {\bibinfo {volume} {2}},\
  \bibinfo {pages} {102301} (\bibinfo {year} {2017})}\BibitemShut {NoStop}%
\bibitem [{\citenamefont {Mandal}\ and\ \citenamefont
  {Khakhar}(2018)}]{mandal2018}%
  \BibitemOpen
  \bibfield  {author} {\bibinfo {author} {\bibfnamefont {S.}~\bibnamefont
  {Mandal}}\ and\ \bibinfo {author} {\bibfnamefont {D.~V.}\ \bibnamefont
  {Khakhar}},\ }\bibfield  {title} {\enquote {\bibinfo {title} {A study of the
  rheology and micro-structure of dumbbells in shear geometries},}\ }\href@noop
  {} {\bibfield  {journal} {\bibinfo  {journal} {Phys. Fluids}\ }\textbf
  {\bibinfo {volume} {30}},\ \bibinfo {pages} {013303} (\bibinfo {year}
  {2018})}\BibitemShut {NoStop}%
\bibitem [{\citenamefont {Bharathraj}\ and\ \citenamefont
  {Kumaran}(2018)}]{bharathraj2018}%
  \BibitemOpen
  \bibfield  {author} {\bibinfo {author} {\bibfnamefont {S.}~\bibnamefont
  {Bharathraj}}\ and\ \bibinfo {author} {\bibfnamefont {V.}~\bibnamefont
  {Kumaran}},\ }\bibfield  {title} {\enquote {\bibinfo {title} {Effect of
  particle stiffness on contact dynamics and rheology in a dense granular
  flow},}\ }\href@noop {} {\bibfield  {journal} {\bibinfo  {journal} {Phys.
  Rev. E}\ }\textbf {\bibinfo {volume} {97}},\ \bibinfo {pages} {012902}
  (\bibinfo {year} {2018})}\BibitemShut {NoStop}%
\bibitem [{\citenamefont {Bhateja}\ and\ \citenamefont
  {Khakhar}(2018)}]{bhateja2018}%
  \BibitemOpen
  \bibfield  {author} {\bibinfo {author} {\bibfnamefont {A.}~\bibnamefont
  {Bhateja}}\ and\ \bibinfo {author} {\bibfnamefont {D.~V.}\ \bibnamefont
  {Khakhar}},\ }\bibfield  {title} {\enquote {\bibinfo {title} {Rheology of
  dense granular flows in two dimensions: Comparison of fully two-dimensional
  flows to unidirectional shear flow},}\ }\href@noop {} {\bibfield  {journal}
  {\bibinfo  {journal} {Phys. Rev. Fluids}\ }\textbf {\bibinfo {volume} {3}},\
  \bibinfo {pages} {062301} (\bibinfo {year} {2018})}\BibitemShut {NoStop}%
\bibitem [{\citenamefont {Berzi}\ and\ \citenamefont
  {Jenkins}(2018)}]{berzi2018}%
  \BibitemOpen
  \bibfield  {author} {\bibinfo {author} {\bibfnamefont {D.}~\bibnamefont
  {Berzi}}\ and\ \bibinfo {author} {\bibfnamefont {J.~T.}\ \bibnamefont
  {Jenkins}},\ }\bibfield  {title} {\enquote {\bibinfo {title} {Fluidity,
  anisotropy, and velocity correlations in frictionless, collisional grain
  flows},}\ }\href@noop {} {\bibfield  {journal} {\bibinfo  {journal} {Phys.
  Rev. Fluids}\ }\textbf {\bibinfo {volume} {3}},\ \bibinfo {pages} {094303}
  (\bibinfo {year} {2018})}\BibitemShut {NoStop}%
\bibitem [{\citenamefont {Mandal}\ and\ \citenamefont
  {Khakhar}(2019)}]{sandip-pof2019}%
  \BibitemOpen
  \bibfield  {author} {\bibinfo {author} {\bibfnamefont {S.}~\bibnamefont
  {Mandal}}\ and\ \bibinfo {author} {\bibfnamefont {D.~V.}\ \bibnamefont
  {Khakhar}},\ }\bibfield  {title} {\enquote {\bibinfo {title} {Dense granular
  flow of mixtures of spheres and dumbbells down a rough inclined plane:
  Segregation and rheology},}\ }\href@noop {} {\bibfield  {journal} {\bibinfo
  {journal} {Phys. of Fluids}\ }\textbf {\bibinfo {volume} {31}},\ \bibinfo
  {pages} {023304} (\bibinfo {year} {2019})}\BibitemShut {NoStop}%
\bibitem [{\citenamefont {Guo}\ and\ \citenamefont {Campbell}(2016)}]{guo2016}%
  \BibitemOpen
  \bibfield  {author} {\bibinfo {author} {\bibfnamefont {T.}~\bibnamefont
  {Guo}}\ and\ \bibinfo {author} {\bibfnamefont {C.~S.}\ \bibnamefont
  {Campbell}},\ }\bibfield  {title} {\enquote {\bibinfo {title} {An
  experimental study of the elastic theory for granular flows},}\ }\href@noop
  {} {\bibfield  {journal} {\bibinfo  {journal} {Phys. of Fluids}\ }\textbf
  {\bibinfo {volume} {28}},\ \bibinfo {pages} {083303} (\bibinfo {year}
  {2016})}\BibitemShut {NoStop}%
\bibitem [{\citenamefont {Staron}, \citenamefont {Lagr{\'e}e},\ and\
  \citenamefont {Popinet}(2012)}]{staron2012}%
  \BibitemOpen
  \bibfield  {author} {\bibinfo {author} {\bibfnamefont {L.}~\bibnamefont
  {Staron}}, \bibinfo {author} {\bibfnamefont {P.-Y.}\ \bibnamefont
  {Lagr{\'e}e}}, \ and\ \bibinfo {author} {\bibfnamefont {S.}~\bibnamefont
  {Popinet}},\ }\bibfield  {title} {\enquote {\bibinfo {title} {The granular
  silo as a continuum plastic flow: The hour-glass vs the clepsydra},}\
  }\href@noop {} {\bibfield  {journal} {\bibinfo  {journal} {Phys. Fluids}\
  }\textbf {\bibinfo {volume} {24}},\ \bibinfo {pages} {103301} (\bibinfo
  {year} {2012})}\BibitemShut {NoStop}%
\bibitem [{\citenamefont {Staron}, \citenamefont {Lagr{\'e}e},\ and\
  \citenamefont {Popinet}(2014)}]{staron2014}%
  \BibitemOpen
  \bibfield  {author} {\bibinfo {author} {\bibfnamefont {L.}~\bibnamefont
  {Staron}}, \bibinfo {author} {\bibfnamefont {P.-Y.}\ \bibnamefont
  {Lagr{\'e}e}}, \ and\ \bibinfo {author} {\bibfnamefont {S.}~\bibnamefont
  {Popinet}},\ }\bibfield  {title} {\enquote {\bibinfo {title} {Continuum
  simulation of the discharge of the granular silo},}\ }\href@noop {}
  {\bibfield  {journal} {\bibinfo  {journal} {Eur. Phys. J. E}\ }\textbf
  {\bibinfo {volume} {37}},\ \bibinfo {pages} {5} (\bibinfo {year}
  {2014})}\BibitemShut {NoStop}%
\bibitem [{\citenamefont {Martin}\ \emph {et~al.}(2017)\citenamefont {Martin},
  \citenamefont {Ionescu}, \citenamefont {Mangeney}, \citenamefont {Bouchut},\
  and\ \citenamefont {Farin}}]{martin2017}%
  \BibitemOpen
  \bibfield  {author} {\bibinfo {author} {\bibfnamefont {N.}~\bibnamefont
  {Martin}}, \bibinfo {author} {\bibfnamefont {I.~R.}\ \bibnamefont {Ionescu}},
  \bibinfo {author} {\bibfnamefont {A.}~\bibnamefont {Mangeney}}, \bibinfo
  {author} {\bibfnamefont {F.}~\bibnamefont {Bouchut}}, \ and\ \bibinfo
  {author} {\bibfnamefont {M.}~\bibnamefont {Farin}},\ }\bibfield  {title}
  {\enquote {\bibinfo {title} {Continuum viscoplastic simulation of a granular
  column collapse on large slopes: $\mu$ ({I}) rheology and lateral wall
  effects},}\ }\href@noop {} {\bibfield  {journal} {\bibinfo  {journal} {Phys.
  Fluids}\ }\textbf {\bibinfo {volume} {29}},\ \bibinfo {pages} {013301}
  (\bibinfo {year} {2017})}\BibitemShut {NoStop}%
\bibitem [{\citenamefont {Luo}, \citenamefont {Zheng},\ and\ \citenamefont
  {Yu}(2019)}]{luo2019}%
  \BibitemOpen
  \bibfield  {author} {\bibinfo {author} {\bibfnamefont {Q.}~\bibnamefont
  {Luo}}, \bibinfo {author} {\bibfnamefont {Q.}~\bibnamefont {Zheng}}, \ and\
  \bibinfo {author} {\bibfnamefont {A.}~\bibnamefont {Yu}},\ }\bibfield
  {title} {\enquote {\bibinfo {title} {Quantitative comparison of hydrodynamic
  and elastoplastic approaches for modeling granular flow in silo},}\
  }\href@noop {} {\bibfield  {journal} {\bibinfo  {journal} {AIChE J.}\
  }\textbf {\bibinfo {volume} {65}} (\bibinfo {year} {2019})}\BibitemShut
  {NoStop}%
\bibitem [{\citenamefont {Kumaran}(2015)}]{kumaran2015}%
  \BibitemOpen
  \bibfield  {author} {\bibinfo {author} {\bibfnamefont {V.}~\bibnamefont
  {Kumaran}},\ }\bibfield  {title} {\enquote {\bibinfo {title} {Kinetic theory
  for sheared granular flows},}\ }\href@noop {} {\bibfield  {journal} {\bibinfo
   {journal} {C. R. Physique}\ }\textbf {\bibinfo {volume} {16}},\ \bibinfo
  {pages} {51--61} (\bibinfo {year} {2015})}\BibitemShut {NoStop}%
\bibitem [{\citenamefont {Jenkins}\ and\ \citenamefont
  {Richman}(1985)}]{jenkins1985}%
  \BibitemOpen
  \bibfield  {author} {\bibinfo {author} {\bibfnamefont {J.~T.}\ \bibnamefont
  {Jenkins}}\ and\ \bibinfo {author} {\bibfnamefont {M.~W.}\ \bibnamefont
  {Richman}},\ }\bibfield  {title} {\enquote {\bibinfo {title} {Kinetic theory
  for plane flows of a dense gas of identical, rough, inelastic, circular
  disks},}\ }\href@noop {} {\bibfield  {journal} {\bibinfo  {journal} {Phys.
  Fluids}\ }\textbf {\bibinfo {volume} {28}},\ \bibinfo {pages} {3485--3494}
  (\bibinfo {year} {1985})}\BibitemShut {NoStop}%
\bibitem [{\citenamefont {Goldhirsch}\ and\ \citenamefont
  {Sela}(1996)}]{goldhirsch1996}%
  \BibitemOpen
  \bibfield  {author} {\bibinfo {author} {\bibfnamefont {I.}~\bibnamefont
  {Goldhirsch}}\ and\ \bibinfo {author} {\bibfnamefont {N.}~\bibnamefont
  {Sela}},\ }\bibfield  {title} {\enquote {\bibinfo {title} {Origin of normal
  stress differences in rapid granular flows},}\ }\href@noop {} {\bibfield
  {journal} {\bibinfo  {journal} {Phys. Rev. E}\ }\textbf {\bibinfo {volume}
  {54}},\ \bibinfo {pages} {4458} (\bibinfo {year} {1996})}\BibitemShut
  {NoStop}%
\bibitem [{\citenamefont {Berzi}\ and\ \citenamefont
  {Vescovi}(2015)}]{berzi2015}%
  \BibitemOpen
  \bibfield  {author} {\bibinfo {author} {\bibfnamefont {D.}~\bibnamefont
  {Berzi}}\ and\ \bibinfo {author} {\bibfnamefont {D.}~\bibnamefont
  {Vescovi}},\ }\bibfield  {title} {\enquote {\bibinfo {title} {Different
  singularities in the functions of extended kinetic theory at the origin of
  the yield stress in granular flows},}\ }\href@noop {} {\bibfield  {journal}
  {\bibinfo  {journal} {Phys. Fluids}\ }\textbf {\bibinfo {volume} {27}},\
  \bibinfo {pages} {013302} (\bibinfo {year} {2015})}\BibitemShut {NoStop}%
\bibitem [{\citenamefont {Saha}\ and\ \citenamefont {Alam}(2016)}]{saha2016}%
  \BibitemOpen
  \bibfield  {author} {\bibinfo {author} {\bibfnamefont {S.}~\bibnamefont
  {Saha}}\ and\ \bibinfo {author} {\bibfnamefont {M.}~\bibnamefont {Alam}},\
  }\bibfield  {title} {\enquote {\bibinfo {title} {Normal stress differences,
  their origin and constitutive relations for a sheared granular fluid},}\
  }\href@noop {} {\bibfield  {journal} {\bibinfo  {journal} {J. Fluid Mech.}\
  }\textbf {\bibinfo {volume} {795}},\ \bibinfo {pages} {549--580} (\bibinfo
  {year} {2016})}\BibitemShut {NoStop}%
\bibitem [{\citenamefont {Duan}\ and\ \citenamefont {Feng}(2019)}]{duan2019}%
  \BibitemOpen
  \bibfield  {author} {\bibinfo {author} {\bibfnamefont {Y.}~\bibnamefont
  {Duan}}\ and\ \bibinfo {author} {\bibfnamefont {Z.-G.}\ \bibnamefont
  {Feng}},\ }\bibfield  {title} {\enquote {\bibinfo {title} {A new kinetic
  theory model of granular flows that incorporates particle stiffness},}\
  }\href@noop {} {\bibfield  {journal} {\bibinfo  {journal} {Phys. Fluids}\
  }\textbf {\bibinfo {volume} {31}},\ \bibinfo {pages} {013301} (\bibinfo
  {year} {2019})}\BibitemShut {NoStop}%
\bibitem [{\citenamefont {Lacaze}\ and\ \citenamefont
  {Kerswell}(2009)}]{lacaze2009}%
  \BibitemOpen
  \bibfield  {author} {\bibinfo {author} {\bibfnamefont {L.}~\bibnamefont
  {Lacaze}}\ and\ \bibinfo {author} {\bibfnamefont {R.~R.}\ \bibnamefont
  {Kerswell}},\ }\bibfield  {title} {\enquote {\bibinfo {title} {Axisymmetric
  granular collapse: {A} transient {3D} flow test of viscoplasticity},}\
  }\href@noop {} {\bibfield  {journal} {\bibinfo  {journal} {Phys. Rev. Lett.}\
  }\textbf {\bibinfo {volume} {102}},\ \bibinfo {pages} {108305} (\bibinfo
  {year} {2009})}\BibitemShut {NoStop}%
\bibitem [{\citenamefont {Yang}\ and\ \citenamefont {Huang}(2016)}]{yang2016}%
  \BibitemOpen
  \bibfield  {author} {\bibinfo {author} {\bibfnamefont {F.~L.}\ \bibnamefont
  {Yang}}\ and\ \bibinfo {author} {\bibfnamefont {Y.~T.}\ \bibnamefont
  {Huang}},\ }\bibfield  {title} {\enquote {\bibinfo {title} {New aspects for
  friction coefficients of finite granular avalanche down a flat narrow
  reservoir},}\ }\href@noop {} {\bibfield  {journal} {\bibinfo  {journal}
  {Granular Matter}\ }\textbf {\bibinfo {volume} {18}},\ \bibinfo {pages} {77}
  (\bibinfo {year} {2016})}\BibitemShut {NoStop}%
\bibitem [{\citenamefont {Saingier}, \citenamefont {Deboeuf},\ and\
  \citenamefont {Lagr{\'e}e}(2016)}]{saingier2016}%
  \BibitemOpen
  \bibfield  {author} {\bibinfo {author} {\bibfnamefont {G.}~\bibnamefont
  {Saingier}}, \bibinfo {author} {\bibfnamefont {S.}~\bibnamefont {Deboeuf}}, \
  and\ \bibinfo {author} {\bibfnamefont {P.-Y.}\ \bibnamefont {Lagr{\'e}e}},\
  }\bibfield  {title} {\enquote {\bibinfo {title} {On the front shape of an
  inertial granular flow down a rough incline},}\ }\href@noop {} {\bibfield
  {journal} {\bibinfo  {journal} {Phys. of Fluids}\ }\textbf {\bibinfo {volume}
  {28}},\ \bibinfo {pages} {053302} (\bibinfo {year} {2016})}\BibitemShut
  {NoStop}%
\bibitem [{\citenamefont {Walton}\ and\ \citenamefont
  {Braun}(1986)}]{walton1986}%
  \BibitemOpen
  \bibfield  {author} {\bibinfo {author} {\bibfnamefont {O.~R.}\ \bibnamefont
  {Walton}}\ and\ \bibinfo {author} {\bibfnamefont {R.~L.}\ \bibnamefont
  {Braun}},\ }\bibfield  {title} {\enquote {\bibinfo {title} {Viscosity,
  granular-temperature, and stress calculations for shearing assemblies of
  inelastic, frictional disks},}\ }\href@noop {} {\bibfield  {journal}
  {\bibinfo  {journal} {J. Rheol}\ }\textbf {\bibinfo {volume} {30}},\ \bibinfo
  {pages} {949--980} (\bibinfo {year} {1986})}\BibitemShut {NoStop}%
\bibitem [{\citenamefont {Campbell}\ and\ \citenamefont
  {Gong}(1986)}]{campbell1986}%
  \BibitemOpen
  \bibfield  {author} {\bibinfo {author} {\bibfnamefont {C.~S.}\ \bibnamefont
  {Campbell}}\ and\ \bibinfo {author} {\bibfnamefont {A.}~\bibnamefont
  {Gong}},\ }\bibfield  {title} {\enquote {\bibinfo {title} {The stress tensor
  in a two-dimensional granular shear flow},}\ }\href@noop {} {\bibfield
  {journal} {\bibinfo  {journal} {J. Fluid Mech.}\ }\textbf {\bibinfo {volume}
  {164}},\ \bibinfo {pages} {107--125} (\bibinfo {year} {1986})}\BibitemShut
  {NoStop}%
\bibitem [{\citenamefont {Jenkins}\ and\ \citenamefont
  {Richman}(1988)}]{jenkins1988}%
  \BibitemOpen
  \bibfield  {author} {\bibinfo {author} {\bibfnamefont {J.~T.}\ \bibnamefont
  {Jenkins}}\ and\ \bibinfo {author} {\bibfnamefont {M.~W.}\ \bibnamefont
  {Richman}},\ }\bibfield  {title} {\enquote {\bibinfo {title} {Plane simple
  shear of smooth inelastic circular disks: the anisotropy of the second moment
  in the dilute and dense limits},}\ }\href@noop {} {\bibfield  {journal}
  {\bibinfo  {journal} {J. Fluid Mech.}\ }\textbf {\bibinfo {volume} {192}},\
  \bibinfo {pages} {313--328} (\bibinfo {year} {1988})}\BibitemShut {NoStop}%
\bibitem [{\citenamefont {Campbell}(1989)}]{campbell1989}%
  \BibitemOpen
  \bibfield  {author} {\bibinfo {author} {\bibfnamefont {C.~S.}\ \bibnamefont
  {Campbell}},\ }\bibfield  {title} {\enquote {\bibinfo {title} {The stress
  tensor for simple shear flows of a granular material},}\ }\href@noop {}
  {\bibfield  {journal} {\bibinfo  {journal} {J. Fluid Mech.}\ }\textbf
  {\bibinfo {volume} {203}},\ \bibinfo {pages} {449--473} (\bibinfo {year}
  {1989})}\BibitemShut {NoStop}%
\bibitem [{\citenamefont {Alam}\ and\ \citenamefont {Luding}(2003)}]{alam2003}%
  \BibitemOpen
  \bibfield  {author} {\bibinfo {author} {\bibfnamefont {M.}~\bibnamefont
  {Alam}}\ and\ \bibinfo {author} {\bibfnamefont {S.}~\bibnamefont {Luding}},\
  }\bibfield  {title} {\enquote {\bibinfo {title} {First normal stress
  difference and crystallization in a dense sheared granular fluid},}\
  }\href@noop {} {\bibfield  {journal} {\bibinfo  {journal} {Phys. Fluids}\
  }\textbf {\bibinfo {volume} {15}},\ \bibinfo {pages} {2298--2312} (\bibinfo
  {year} {2003})}\BibitemShut {NoStop}%
\bibitem [{\citenamefont {Alam}\ and\ \citenamefont {Luding}(2005)}]{alam2005}%
  \BibitemOpen
  \bibfield  {author} {\bibinfo {author} {\bibfnamefont {M.}~\bibnamefont
  {Alam}}\ and\ \bibinfo {author} {\bibfnamefont {S.}~\bibnamefont {Luding}},\
  }\bibfield  {title} {\enquote {\bibinfo {title} {Non-newtonian granular
  fluid: {S}imulation and theory},}\ }in\ \href@noop {} {\emph {\bibinfo
  {booktitle} {Powders and Grains}}},\ \bibinfo {editor} {edited by\ \bibinfo
  {editor} {\bibfnamefont {R.}~\bibnamefont {Garcia-Rojo}}, \bibinfo {editor}
  {\bibfnamefont {H.~J.}\ \bibnamefont {Herrmann}}, \ and\ \bibinfo {editor}
  {\bibfnamefont {S.}~\bibnamefont {McNamara}}}\ (\bibinfo {year} {2005})\ pp.\
  \bibinfo {pages} {1141--1144}\BibitemShut {NoStop}%
\bibitem [{\citenamefont {Zuriguel}\ \emph {et~al.}(2005)\citenamefont
  {Zuriguel}, \citenamefont {Garcimart{\'\i}n}, \citenamefont {Maza},
  \citenamefont {Pugnaloni},\ and\ \citenamefont {Pastor}}]{zuriguel2005}%
  \BibitemOpen
  \bibfield  {author} {\bibinfo {author} {\bibfnamefont {I.}~\bibnamefont
  {Zuriguel}}, \bibinfo {author} {\bibfnamefont {A.}~\bibnamefont
  {Garcimart{\'\i}n}}, \bibinfo {author} {\bibfnamefont {D.}~\bibnamefont
  {Maza}}, \bibinfo {author} {\bibfnamefont {L.~A.}\ \bibnamefont {Pugnaloni}},
  \ and\ \bibinfo {author} {\bibfnamefont {J.~M.}\ \bibnamefont {Pastor}},\
  }\bibfield  {title} {\enquote {\bibinfo {title} {Jamming during the discharge
  of granular matter from a silo},}\ }\href@noop {} {\bibfield  {journal}
  {\bibinfo  {journal} {Phys. Rev. E}\ }\textbf {\bibinfo {volume} {71}},\
  \bibinfo {pages} {051303} (\bibinfo {year} {2005})}\BibitemShut {NoStop}%
\bibitem [{\citenamefont {Mankoc}\ \emph {et~al.}(2007)\citenamefont {Mankoc},
  \citenamefont {Janda}, \citenamefont {Arevalo}, \citenamefont {Pastor},
  \citenamefont {Zuriguel}, \citenamefont {Garcimart{\'\i}n},\ and\
  \citenamefont {Maza}}]{mankoc2007}%
  \BibitemOpen
  \bibfield  {author} {\bibinfo {author} {\bibfnamefont {C.}~\bibnamefont
  {Mankoc}}, \bibinfo {author} {\bibfnamefont {A.}~\bibnamefont {Janda}},
  \bibinfo {author} {\bibfnamefont {R.}~\bibnamefont {Arevalo}}, \bibinfo
  {author} {\bibfnamefont {J.~M.}\ \bibnamefont {Pastor}}, \bibinfo {author}
  {\bibfnamefont {I.}~\bibnamefont {Zuriguel}}, \bibinfo {author}
  {\bibfnamefont {A.}~\bibnamefont {Garcimart{\'\i}n}}, \ and\ \bibinfo
  {author} {\bibfnamefont {D.}~\bibnamefont {Maza}},\ }\bibfield  {title}
  {\enquote {\bibinfo {title} {The flow rate of granular materials through an
  orifice},}\ }\href@noop {} {\bibfield  {journal} {\bibinfo  {journal}
  {Granular Matter}\ }\textbf {\bibinfo {volume} {9}},\ \bibinfo {pages}
  {407--414} (\bibinfo {year} {2007})}\BibitemShut {NoStop}%
\bibitem [{\citenamefont {Sh{\"a}fer}, \citenamefont {Dippel},\ and\
  \citenamefont {Wolf}(1996)}]{shafer1996}%
  \BibitemOpen
  \bibfield  {author} {\bibinfo {author} {\bibfnamefont {J.}~\bibnamefont
  {Sh{\"a}fer}}, \bibinfo {author} {\bibfnamefont {S.}~\bibnamefont {Dippel}},
  \ and\ \bibinfo {author} {\bibfnamefont {D.~E.}\ \bibnamefont {Wolf}},\
  }\bibfield  {title} {\enquote {\bibinfo {title} {Force schemes in simulations
  of granular materials},}\ }\href@noop {} {\bibfield  {journal} {\bibinfo
  {journal} {J. Phys. I}\ }\textbf {\bibinfo {volume} {6}},\ \bibinfo {pages}
  {5--20} (\bibinfo {year} {1996})}\BibitemShut {NoStop}%
\bibitem [{\citenamefont {Zhang}\ and\ \citenamefont
  {Whiten}(1996)}]{zhang1996}%
  \BibitemOpen
  \bibfield  {author} {\bibinfo {author} {\bibfnamefont {D.}~\bibnamefont
  {Zhang}}\ and\ \bibinfo {author} {\bibfnamefont {W.~J.}\ \bibnamefont
  {Whiten}},\ }\bibfield  {title} {\enquote {\bibinfo {title} {The calculation
  of contact forces between particles using spring and damping models},}\
  }\href@noop {} {\bibfield  {journal} {\bibinfo  {journal} {Powder
  Technology}\ }\textbf {\bibinfo {volume} {88}},\ \bibinfo {pages} {59--64}
  (\bibinfo {year} {1996})}\BibitemShut {NoStop}%
\bibitem [{\citenamefont {Stevens}\ and\ \citenamefont
  {Hrenya}(2005)}]{stevens2005}%
  \BibitemOpen
  \bibfield  {author} {\bibinfo {author} {\bibfnamefont {A.~B.}\ \bibnamefont
  {Stevens}}\ and\ \bibinfo {author} {\bibfnamefont {C.~M.}\ \bibnamefont
  {Hrenya}},\ }\bibfield  {title} {\enquote {\bibinfo {title} {Comparison of
  soft-sphere models to measurements of collision properties during normal
  impacts},}\ }\href@noop {} {\bibfield  {journal} {\bibinfo  {journal} {Powder
  Technology}\ }\textbf {\bibinfo {volume} {154}},\ \bibinfo {pages} {99--109}
  (\bibinfo {year} {2005})}\BibitemShut {NoStop}%
\bibitem [{\citenamefont {Kruggel-Emden}, \citenamefont {Wirtz},\ and\
  \citenamefont {Scherer}(2008)}]{kruggel2008a}%
  \BibitemOpen
  \bibfield  {author} {\bibinfo {author} {\bibfnamefont {H.}~\bibnamefont
  {Kruggel-Emden}}, \bibinfo {author} {\bibfnamefont {S.}~\bibnamefont
  {Wirtz}}, \ and\ \bibinfo {author} {\bibfnamefont {V.}~\bibnamefont
  {Scherer}},\ }\bibfield  {title} {\enquote {\bibinfo {title} {A study on
  tangential force laws applicable to the discrete element method (dem) for
  materials with viscoelastic or plastic behavior},}\ }\href@noop {} {\bibfield
   {journal} {\bibinfo  {journal} {Chem. Eng. Sci.}\ }\textbf {\bibinfo
  {volume} {63}},\ \bibinfo {pages} {1523--1541} (\bibinfo {year}
  {2008})}\BibitemShut {NoStop}%
\bibitem [{\citenamefont {Kruggel-Emden}\ \emph {et~al.}(2007)\citenamefont
  {Kruggel-Emden}, \citenamefont {Simsek}, \citenamefont {Rickelt},
  \citenamefont {Wirtz},\ and\ \citenamefont {Scherer}}]{kruggel2007}%
  \BibitemOpen
  \bibfield  {author} {\bibinfo {author} {\bibfnamefont {H.}~\bibnamefont
  {Kruggel-Emden}}, \bibinfo {author} {\bibfnamefont {E.}~\bibnamefont
  {Simsek}}, \bibinfo {author} {\bibfnamefont {S.}~\bibnamefont {Rickelt}},
  \bibinfo {author} {\bibfnamefont {S.}~\bibnamefont {Wirtz}}, \ and\ \bibinfo
  {author} {\bibfnamefont {V.}~\bibnamefont {Scherer}},\ }\bibfield  {title}
  {\enquote {\bibinfo {title} {Review and extension of normal force models for
  the discrete element method},}\ }\href@noop {} {\bibfield  {journal}
  {\bibinfo  {journal} {Powder Technology}\ }\textbf {\bibinfo {volume}
  {171}},\ \bibinfo {pages} {157--173} (\bibinfo {year} {2007})}\BibitemShut
  {NoStop}%
\bibitem [{\citenamefont {Allen}\ and\ \citenamefont
  {Tildesley}(1989)}]{allen1989}%
  \BibitemOpen
  \bibfield  {author} {\bibinfo {author} {\bibfnamefont {M.~P.}\ \bibnamefont
  {Allen}}\ and\ \bibinfo {author} {\bibfnamefont {D.~J.}\ \bibnamefont
  {Tildesley}},\ }\href@noop {} {\emph {\bibinfo {title} {Computer simulation
  of liquids}}}\ (\bibinfo {year} {1989})\BibitemShut {NoStop}%
\bibitem [{\citenamefont {Kruggel-Emden}\ \emph {et~al.}(2008)\citenamefont
  {Kruggel-Emden}, \citenamefont {Sturm}, \citenamefont {Wirtz},\ and\
  \citenamefont {Scherer}}]{kruggel2008b}%
  \BibitemOpen
  \bibfield  {author} {\bibinfo {author} {\bibfnamefont {H.}~\bibnamefont
  {Kruggel-Emden}}, \bibinfo {author} {\bibfnamefont {M.}~\bibnamefont
  {Sturm}}, \bibinfo {author} {\bibfnamefont {S.}~\bibnamefont {Wirtz}}, \ and\
  \bibinfo {author} {\bibfnamefont {V.}~\bibnamefont {Scherer}},\ }\bibfield
  {title} {\enquote {\bibinfo {title} {Selection of an appropriate time
  integration scheme for the discrete element method ({DEM})},}\ }\href@noop {}
  {\bibfield  {journal} {\bibinfo  {journal} {Comput. Chem. Eng.}\ }\textbf
  {\bibinfo {volume} {32}},\ \bibinfo {pages} {2263--2279} (\bibinfo {year}
  {2008})}\BibitemShut {NoStop}%
\bibitem [{\citenamefont {Weinhart}\ \emph {et~al.}(2013)\citenamefont
  {Weinhart}, \citenamefont {Hartkamp}, \citenamefont {Thornton},\ and\
  \citenamefont {Luding}}]{weinhart2013}%
  \BibitemOpen
  \bibfield  {author} {\bibinfo {author} {\bibfnamefont {T.}~\bibnamefont
  {Weinhart}}, \bibinfo {author} {\bibfnamefont {R.}~\bibnamefont {Hartkamp}},
  \bibinfo {author} {\bibfnamefont {A.~R.}\ \bibnamefont {Thornton}}, \ and\
  \bibinfo {author} {\bibfnamefont {S.}~\bibnamefont {Luding}},\ }\bibfield
  {title} {\enquote {\bibinfo {title} {Coarse-grained local and objective
  continuum description of three-dimensional granular flows down an inclined
  surface},}\ }\href@noop {} {\bibfield  {journal} {\bibinfo  {journal} {Phys.
  Fluids}\ }\textbf {\bibinfo {volume} {25}},\ \bibinfo {pages} {070605}
  (\bibinfo {year} {2013})}\BibitemShut {NoStop}%
\bibitem [{\citenamefont {Artoni}\ and\ \citenamefont
  {Richard}(2015)}]{artoni2015}%
  \BibitemOpen
  \bibfield  {author} {\bibinfo {author} {\bibfnamefont {R.}~\bibnamefont
  {Artoni}}\ and\ \bibinfo {author} {\bibfnamefont {P.}~\bibnamefont
  {Richard}},\ }\bibfield  {title} {\enquote {\bibinfo {title} {Average balance
  equations, scale dependence, and energy cascade for granular materials},}\
  }\href@noop {} {\bibfield  {journal} {\bibinfo  {journal} {Phys. Rev. E}\
  }\textbf {\bibinfo {volume} {91}},\ \bibinfo {pages} {032202} (\bibinfo
  {year} {2015})}\BibitemShut {NoStop}%
\bibitem [{\citenamefont {Tripathi}\ and\ \citenamefont
  {Khakhar}(2010)}]{tripathi2010}%
  \BibitemOpen
  \bibfield  {author} {\bibinfo {author} {\bibfnamefont {A.}~\bibnamefont
  {Tripathi}}\ and\ \bibinfo {author} {\bibfnamefont {D.~V.}\ \bibnamefont
  {Khakhar}},\ }\bibfield  {title} {\enquote {\bibinfo {title} {Steady flow of
  smooth, inelastic particles on a bumpy inclined plane: Hard and soft particle
  simulations},}\ }\href@noop {} {\bibfield  {journal} {\bibinfo  {journal}
  {Phys. Rev. E}\ }\textbf {\bibinfo {volume} {81}},\ \bibinfo {pages} {041307}
  (\bibinfo {year} {2010})}\BibitemShut {NoStop}%
\bibitem [{\citenamefont {Altman}\ and\ \citenamefont
  {Bland}(2005)}]{altman2005}%
  \BibitemOpen
  \bibfield  {author} {\bibinfo {author} {\bibfnamefont {D.~G.}\ \bibnamefont
  {Altman}}\ and\ \bibinfo {author} {\bibfnamefont {J.~M.}\ \bibnamefont
  {Bland}},\ }\bibfield  {title} {\enquote {\bibinfo {title} {Standard
  deviations and standard errors},}\ }\href@noop {} {\bibfield  {journal}
  {\bibinfo  {journal} {BMJ}\ }\textbf {\bibinfo {volume} {331}},\ \bibinfo
  {pages} {903} (\bibinfo {year} {2005})}\BibitemShut {NoStop}%
\bibitem [{\citenamefont {Rycroft}, \citenamefont {Kamrin},\ and\ \citenamefont
  {Bazant}(2009)}]{rycroft2009}%
  \BibitemOpen
  \bibfield  {author} {\bibinfo {author} {\bibfnamefont {C.~H.}\ \bibnamefont
  {Rycroft}}, \bibinfo {author} {\bibfnamefont {K.}~\bibnamefont {Kamrin}}, \
  and\ \bibinfo {author} {\bibfnamefont {M.~Z.}\ \bibnamefont {Bazant}},\
  }\bibfield  {title} {\enquote {\bibinfo {title} {Assessing continuum
  postulates in simulations of granular flow},}\ }\href@noop {} {\bibfield
  {journal} {\bibinfo  {journal} {J. Mech. Phys. Solids}\ }\textbf {\bibinfo
  {volume} {57}},\ \bibinfo {pages} {828--839} (\bibinfo {year}
  {2009})}\BibitemShut {NoStop}%
\bibitem [{\citenamefont {Rao}\ and\ \citenamefont {Nott}(2008)}]{rao_nott}%
  \BibitemOpen
  \bibfield  {author} {\bibinfo {author} {\bibfnamefont {K.}~\bibnamefont
  {Rao}}\ and\ \bibinfo {author} {\bibfnamefont {P.}~\bibnamefont {Nott}},\
  }\href@noop {} {\emph {\bibinfo {title} {An {I}ntroduction to {G}ranular
  {F}low}}}\ (\bibinfo  {publisher} {Cambridge {U}niversity {P}ress},\ \bibinfo
  {year} {2008})\BibitemShut {NoStop}%
\bibitem [{\citenamefont {Wagner}\ and\ \citenamefont
  {McKinley}(2016)}]{wagner2016}%
  \BibitemOpen
  \bibfield  {author} {\bibinfo {author} {\bibfnamefont {C.~E.}\ \bibnamefont
  {Wagner}}\ and\ \bibinfo {author} {\bibfnamefont {G.~H.}\ \bibnamefont
  {McKinley}},\ }\bibfield  {title} {\enquote {\bibinfo {title} {The importance
  of flow history in mixed shear and extensional flows},}\ }\href@noop {}
  {\bibfield  {journal} {\bibinfo  {journal} {J. Non-Newtonian Fluid Mech.}\
  }\textbf {\bibinfo {volume} {233}},\ \bibinfo {pages} {133--145} (\bibinfo
  {year} {2016})}\BibitemShut {NoStop}%
\bibitem [{\citenamefont {Lee}\ \emph {et~al.}(2007)\citenamefont {Lee},
  \citenamefont {Dylla-Spears}, \citenamefont {Teclemariam},\ and\
  \citenamefont {Muller}}]{lee2007}%
  \BibitemOpen
  \bibfield  {author} {\bibinfo {author} {\bibfnamefont {J.~S.}\ \bibnamefont
  {Lee}}, \bibinfo {author} {\bibfnamefont {R.}~\bibnamefont {Dylla-Spears}},
  \bibinfo {author} {\bibfnamefont {N.~P.}\ \bibnamefont {Teclemariam}}, \ and\
  \bibinfo {author} {\bibfnamefont {S.~J.}\ \bibnamefont {Muller}},\ }\bibfield
   {title} {\enquote {\bibinfo {title} {Microfluidic four-roll mill for all
  flow types},}\ }\href@noop {} {\bibfield  {journal} {\bibinfo  {journal}
  {Appl. Phys. Lett.}\ }\textbf {\bibinfo {volume} {90}},\ \bibinfo {pages}
  {074103} (\bibinfo {year} {2007})}\BibitemShut {NoStop}%
\bibitem [{\citenamefont {Hudson}\ \emph {et~al.}(2004)\citenamefont {Hudson},
  \citenamefont {Phelan~Jr.}, \citenamefont {Handler}, \citenamefont {Cabral},
  \citenamefont {Migler},\ and\ \citenamefont {Amis}}]{hudson2004}%
  \BibitemOpen
  \bibfield  {author} {\bibinfo {author} {\bibfnamefont {S.~D.}\ \bibnamefont
  {Hudson}}, \bibinfo {author} {\bibfnamefont {F.~R.}\ \bibnamefont
  {Phelan~Jr.}}, \bibinfo {author} {\bibfnamefont {M.~D.}\ \bibnamefont
  {Handler}}, \bibinfo {author} {\bibfnamefont {J.~T.}\ \bibnamefont {Cabral}},
  \bibinfo {author} {\bibfnamefont {K.~B.}\ \bibnamefont {Migler}}, \ and\
  \bibinfo {author} {\bibfnamefont {E.~J.}\ \bibnamefont {Amis}},\ }\bibfield
  {title} {\enquote {\bibinfo {title} {Microfluidic analog of the four-roll
  mill},}\ }\href@noop {} {\bibfield  {journal} {\bibinfo  {journal} {Appl.
  Phys. Lett.}\ }\textbf {\bibinfo {volume} {85}},\ \bibinfo {pages} {335--337}
  (\bibinfo {year} {2004})}\BibitemShut {NoStop}%
\bibitem [{\citenamefont {Kumaran}\ and\ \citenamefont
  {Maheshwari}(2012)}]{kumaran2012}%
  \BibitemOpen
  \bibfield  {author} {\bibinfo {author} {\bibfnamefont {V.}~\bibnamefont
  {Kumaran}}\ and\ \bibinfo {author} {\bibfnamefont {S.}~\bibnamefont
  {Maheshwari}},\ }\bibfield  {title} {\enquote {\bibinfo {title} {Transition
  due to base roughness in a dense granular flow down an inclined plane},}\
  }\href@noop {} {\bibfield  {journal} {\bibinfo  {journal} {Phys. Fluids}\
  }\textbf {\bibinfo {volume} {24}},\ \bibinfo {pages} {053302} (\bibinfo
  {year} {2012})}\BibitemShut {NoStop}%
\bibitem [{\citenamefont {Campbell}(2006)}]{campbell2006}%
  \BibitemOpen
  \bibfield  {author} {\bibinfo {author} {\bibfnamefont {C.~S.}\ \bibnamefont
  {Campbell}},\ }\bibfield  {title} {\enquote {\bibinfo {title} {Granular
  material flows--an overview},}\ }\href@noop {} {\bibfield  {journal}
  {\bibinfo  {journal} {Powder Technology}\ }\textbf {\bibinfo {volume}
  {162}},\ \bibinfo {pages} {208--229} (\bibinfo {year} {2006})}\BibitemShut
  {NoStop}%
\bibitem [{\citenamefont {Hatano}(2007)}]{hatano2007}%
  \BibitemOpen
  \bibfield  {author} {\bibinfo {author} {\bibfnamefont {T.}~\bibnamefont
  {Hatano}},\ }\bibfield  {title} {\enquote {\bibinfo {title} {Power-law
  friction in closely packed granular materials},}\ }\href@noop {} {\bibfield
  {journal} {\bibinfo  {journal} {Phys. Rev. E}\ }\textbf {\bibinfo {volume}
  {75}},\ \bibinfo {pages} {060301} (\bibinfo {year} {2007})}\BibitemShut
  {NoStop}%
\end{thebibliography}%
\end{document}